\documentclass[twocolumn,tighten,twocolappendix]{aastex63}
\usepackage{graphicx}\graphicspath{{./}{./figures/}{./tex/figures}}
\usepackage{multirow}
\usepackage{listings}
\usepackage{enumitem}
\usepackage{hyperref}

\usepackage{soul} 
\usepackage{amsmath}
\usepackage{amssymb}
\usepackage{xspace}
\usepackage{xifthen}

\mathchardef\mhyphen="2D

\newlength{\dhatheight}

\newcommand{\code}[1]{\texttt{#1}\xspace}

\newcommand{\unit}[1]{\ensuremath{\mathrm{\,#1}}\xspace}
\newcommand{\Gyr}{\unit{Gyr}}

\newcommand{\km}{\unit{km}}
\newcommand{\kms}{\km \second^{-1}}
\newcommand{\pc}{\unit{pc}}
\newcommand{\kpc}{\unit{kpc}}
\newcommand{\Mpc}{\unit{Mpc}}
\newcommand{\second}{\unit{s}}

\newcommand{\Msun}{\unit{M_\odot}}

\newcommand{\e}{\unit{e^{-}}}

\newcommand{\h}{\ensuremath{\,h}\xspace}
\newcommand{\dex}{\unit{dex}}

\newcommand{\secref}[1]{Section~\ref{sec:#1}}

\newcommand{\tabref}[1]{Table~\ref{tab:#1}}

\newcommand{\figref}[1]{Figure~\ref{fig:#1}}

\newcommand{\eqnref}[1]{Equation~\eqref{eq:#1}}

\newcommand{\bandvar}[2][]{%
  \ifthenelse{\isempty{#1}}{\var{#2}}{\var{#2\_#1}}%
}

\newcommand{\emcee}{\code{emcee}}

\newcommand{\var}[1]{\ensuremath{\texttt{\MakeUppercase{#1}}}\xspace}

\providecommand\physrep{\ref@jnl{Phys.~Rep.}}%
\providecommand\apjs{\ref@jnl{ApJS}}%
\providecommand{\jcap}{\ref@jnl{JCAP}}%

\shorttitle{Milky Way Satellite Census. IV.}
\shortauthors{Mau \& Nadler et al.}

\begin{document}

\title{Milky Way Satellite Census. IV. Constraints on Decaying Dark Matter from Observations of Milky Way Satellite Galaxies}


\author[0000-0003-3519-4004 ]{S.~Mau}
\affiliation{Department of Physics, Stanford University, 382 Via Pueblo Mall, Stanford, CA 94305, USA}
\affiliation{Kavli Institute for Particle Astrophysics \& Cosmology, P. O. Box 2450, Stanford University, Stanford, CA 94305, USA}

\author[0000-0002-1182-3825 ]{E.~O.~Nadler}
\affiliation{Carnegie Observatories, 813 Santa Barbara Street, Pasadena, CA 91101, USA}
\affiliation{Department of Physics $\&$ Astronomy, University of Southern California, Los Angeles, CA, 90007, USA}

\author[0000-0003-2229-011X]{R.~H.~Wechsler}
\affiliation{Department of Physics, Stanford University, 382 Via Pueblo Mall, Stanford, CA 94305, USA}
\affiliation{Kavli Institute for Particle Astrophysics \& Cosmology, P. O. Box 2450, Stanford University, Stanford, CA 94305, USA}
\affiliation{SLAC National Accelerator Laboratory, Menlo Park, CA 94025, USA}

\author[0000-0001-8251-933X]{A.~Drlica-Wagner}
\affiliation{Department of Astronomy and Astrophysics, University of Chicago, Chicago, IL 60637, USA}
\affiliation{Fermi National Accelerator Laboratory, P. O. Box 500, Batavia, IL 60510, USA}
\affiliation{Kavli Institute for Cosmological Physics, University of Chicago, Chicago, IL 60637, USA}

\author[0000-0001-8156-0429]{K.~Bechtol}
\affiliation{Physics Department, 2320 Chamberlin Hall, University of Wisconsin-Madison, 1150 University Avenue Madison, WI  53706-1390}

\author[0000-0001-5417-2260]{G.~Green}
\affiliation{Max Planck Institute for Astronomy, K\"onigstuhl 17, D-69117, Heidelberg, Germany}

\author[0000-0001-6558-0112]{D.~Huterer}
\affiliation{Department of Physics, University of Michigan, Ann Arbor, MI 48109, USA}

\author[0000-0002-9110-6163]{T.~S.~Li}
\affiliation{Department of Astronomy and Astrophysics, University of Toronto, 50 St. George Street, Toronto ON, M5S 3H4, Canada}

\author[0000-0002-1200-0820]{Y.-Y.~Mao}
\altaffiliation{NHFP Einstein Fellow}
\affiliation{Department of Physics and Astronomy, Rutgers, The State University of New Jersey, Piscataway, NJ 08854, USA}

\author[0000-0002-9144-7726]{C.~E.~Mart\'inez-V\'azquez}
\affiliation{Cerro Tololo Inter-American Observatory, NSF's National Optical-Infrared Astronomy Research Laboratory, Casilla 603, La Serena, Chile}

\author[0000-0001-5435-7820]{M.~McNanna}
\affiliation{Physics Department, 2320 Chamberlin Hall, University of Wisconsin-Madison, 1150 University Avenue Madison, WI  53706-1390}

\author[0000-0001-9649-4815]{B.~Mutlu-Pakdil}
\affiliation{Kavli Institute for Cosmological Physics, University of Chicago, Chicago, IL 60637, USA}

\author[0000-0002-6021-8760]{A.~B.~Pace}
\affiliation{Department of Physics, Carnegie Mellon University, Pittsburgh, Pennsylvania 15312, USA}

\author[0000-0002-8040-6785]{A.~Peter}
\affiliation{Center for Cosmology and Astro-Particle Physics, The Ohio State University, Columbus, OH 43210, USA}

\author[0000-0001-5805-5766]{A.~H.~Riley}
\affiliation{George P. and Cynthia Woods Mitchell Institute for Fundamental Physics and Astronomy, and Department of Physics and Astronomy, Texas A\&M University, College Station, TX 77843,  USA}

\author[0000-0001-5672-6079]{L.~Strigari}
\affiliation{George P. and Cynthia Woods Mitchell Institute for Fundamental Physics and Astronomy, and Department of Physics and Astronomy, Texas A\&M University, College Station, TX 77843,  USA}

\author[0000-0002-8226-6237]{M.-Y.~Wang}
\affiliation{Department of Physics, Carnegie Mellon University, Pittsburgh, Pennsylvania 15312, USA}

\author{M.~Aguena}
\affiliation{Laborat\'orio Interinstitucional de e-Astronomia - LIneA, Rua Gal. Jos\'e Cristino 77, Rio de Janeiro, RJ - 20921-400, Brazil}

\author[0000-0002-7069-7857]{S.~Allam}
\affiliation{Fermi National Accelerator Laboratory, P. O. Box 500, Batavia, IL 60510, USA}

\author[0000-0002-0609-3987]{J.~Annis}
\affiliation{Fermi National Accelerator Laboratory, P. O. Box 500, Batavia, IL 60510, USA}

\author{D.~Bacon}
\affiliation{Institute of Cosmology and Gravitation, University of Portsmouth, Portsmouth, PO1 3FX, UK}

\author{E.~Bertin}
\affiliation{CNRS, UMR 7095, Institut d'Astrophysique de Paris, F-75014, Paris, France}
\affiliation{Sorbonne Universit\'es, UPMC Univ Paris 06, UMR 7095, Institut d'Astrophysique de Paris, F-75014, Paris, France}

\author[0000-0002-4900-805X]{S.~Bocquet}
\affiliation{Faculty of Physics, Ludwig-Maximilians-Universit\"at, Scheinerstr. 1, 81679 Munich, Germany}

\author[0000-0002-8458-5047]{D.~Brooks}
\affiliation{Department of Physics \& Astronomy, University College London, Gower Street, London, WC1E 6BT, UK}

\author{D.~L.~Burke}
\affiliation{Kavli Institute for Particle Astrophysics \& Cosmology, P. O. Box 2450, Stanford University, Stanford, CA 94305, USA}
\affiliation{SLAC National Accelerator Laboratory, Menlo Park, CA 94025, USA}

\author[0000-0003-3044-5150]{A.~Carnero~Rosell}
\affiliation{Instituto de Astrofisica de Canarias, E-38205 La Laguna, Tenerife, Spain}
\affiliation{Laborat\'orio Interinstitucional de e-Astronomia - LIneA, Rua Gal. Jos\'e Cristino 77, Rio de Janeiro, RJ - 20921-400, Brazil}
\affiliation{Universidad de La Laguna, Dpto. Astrofísica, E-38206 La Laguna, Tenerife, Spain}

\author[0000-0002-4802-3194]{M.~Carrasco~Kind}
\affiliation{Center for Astrophysical Surveys, National Center for Supercomputing Applications, 1205 West Clark St., Urbana, IL 61801, USA}
\affiliation{Department of Astronomy, University of Illinois at Urbana-Champaign, 1002 W. Green Street, Urbana, IL 61801, USA}

\author[0000-0002-3130-0204]{J.~Carretero}
\affiliation{Institut de F\'{\i}sica d'Altes Energies (IFAE), The Barcelona Institute of Science and Technology, Campus UAB, 08193 Bellaterra (Barcelona) Spain}

\author{M.~Costanzi}
\affiliation{Astronomy Unit, Department of Physics, University of Trieste, via Tiepolo 11, I-34131 Trieste, Italy}
\affiliation{INAF-Osservatorio Astronomico di Trieste, via G. B. Tiepolo 11, I-34143 Trieste, Italy}
\affiliation{Institute for Fundamental Physics of the Universe, Via Beirut 2, 34014 Trieste, Italy}

\author[0000-0002-9745-6228]{M.~Crocce}
\affiliation{Institut d'Estudis Espacials de Catalunya (IEEC), 08034 Barcelona, Spain}
\affiliation{Institute of Space Sciences (ICE, CSIC),  Campus UAB, Carrer de Can Magrans, s/n,  08193 Barcelona, Spain}

\author{M.~E.~S.~Pereira}
\affiliation{Department of Physics, University of Michigan, Ann Arbor, MI 48109, USA}
\affiliation{Hamburger Sternwarte, Universit\"{a}t Hamburg, Gojenbergsweg 112, 21029 Hamburg, Germany}

\author[0000-0002-4213-8783]{T.~M.~Davis}
\affiliation{School of Mathematics and Physics, University of Queensland,  Brisbane, QLD 4072, Australia}

\author[0000-0001-8318-6813]{J.~De~Vicente}
\affiliation{Centro de Investigaciones Energ\'eticas, Medioambientales y Tecnol\'ogicas (CIEMAT), Madrid, Spain}

\author[0000-0002-0466-3288]{S.~Desai}
\affiliation{Department of Physics, IIT Hyderabad, Kandi, Telangana 502285, India}

\author{P.~Doel}
\affiliation{Department of Physics \& Astronomy, University College London, Gower Street, London, WC1E 6BT, UK}

\author{I.~Ferrero}
\affiliation{Institute of Theoretical Astrophysics, University of Oslo. P.O. Box 1029 Blindern, NO-0315 Oslo, Norway}

\author[0000-0002-2367-5049]{B.~Flaugher}
\affiliation{Fermi National Accelerator Laboratory, P. O. Box 500, Batavia, IL 60510, USA}

\author[0000-0003-4079-3263]{J.~Frieman}
\affiliation{Fermi National Accelerator Laboratory, P. O. Box 500, Batavia, IL 60510, USA}
\affiliation{Kavli Institute for Cosmological Physics, University of Chicago, Chicago, IL 60637, USA}

\author[0000-0002-9370-8360]{J.~Garc\'ia-Bellido}
\affiliation{Instituto de Fisica Teorica UAM/CSIC, Universidad Autonoma de Madrid, 28049 Madrid, Spain}

\author{M.~Gatti}
\affiliation{Department of Physics and Astronomy, University of Pennsylvania, Philadelphia, PA 19104, USA}

\author[0000-0002-3730-1750]{G.~Giannini}
\affiliation{Institut de F\'{\i}sica d'Altes Energies (IFAE), The Barcelona Institute of Science and Technology, Campus UAB, 08193 Bellaterra (Barcelona) Spain}

\author[0000-0003-3270-7644]{D.~Gruen}
\affiliation{Excellence Cluster Origins, Boltzmannstr.\ 2, 85748 Garching, Germany}
\affiliation{Faculty of Physics, Ludwig-Maximilians-Universit\"at, Scheinerstr. 1, 81679 Munich, Germany}

\author{R.~A.~Gruendl}
\affiliation{Center for Astrophysical Surveys, National Center for Supercomputing Applications, 1205 West Clark St., Urbana, IL 61801, USA}
\affiliation{Department of Astronomy, University of Illinois at Urbana-Champaign, 1002 W. Green Street, Urbana, IL 61801, USA}

\author[0000-0003-3023-8362]{J.~Gschwend}
\affiliation{Laborat\'orio Interinstitucional de e-Astronomia - LIneA, Rua Gal. Jos\'e Cristino 77, Rio de Janeiro, RJ - 20921-400, Brazil}
\affiliation{Observat\'orio Nacional, Rua Gal. Jos\'e Cristino 77, Rio de Janeiro, RJ - 20921-400, Brazil}

\author[0000-0003-0825-0517]{G.~Gutierrez}
\affiliation{Fermi National Accelerator Laboratory, P. O. Box 500, Batavia, IL 60510, USA}

\author{S.~R.~Hinton}
\affiliation{School of Mathematics and Physics, University of Queensland,  Brisbane, QLD 4072, Australia}

\author{D.~L.~Hollowood}
\affiliation{Santa Cruz Institute for Particle Physics, Santa Cruz, CA 95064, USA}

\author[0000-0002-6550-2023]{K.~Honscheid}
\affiliation{Center for Cosmology and Astro-Particle Physics, The Ohio State University, Columbus, OH 43210, USA}
\affiliation{Department of Physics, The Ohio State University, Columbus, OH 43210, USA}

\author[0000-0001-5160-4486]{D.~J.~James}
\affiliation{Center for Astrophysics $\vert$ Harvard \& Smithsonian, 60 Garden Street, Cambridge, MA 02138, USA}

\author[0000-0003-0120-0808]{K.~Kuehn}
\affiliation{Australian Astronomical Optics, Macquarie University, North Ryde, NSW 2113, Australia}
\affiliation{Lowell Observatory, 1400 Mars Hill Rd, Flagstaff, AZ 86001, USA}

\author[0000-0002-1134-9035]{O.~Lahav}
\affiliation{Department of Physics \& Astronomy, University College London, Gower Street, London, WC1E 6BT, UK}

\author[0000-0001-9856-9307]{M.~A.~G.~Maia}
\affiliation{Laborat\'orio Interinstitucional de e-Astronomia - LIneA, Rua Gal. Jos\'e Cristino 77, Rio de Janeiro, RJ - 20921-400, Brazil}
\affiliation{Observat\'orio Nacional, Rua Gal. Jos\'e Cristino 77, Rio de Janeiro, RJ - 20921-400, Brazil}

\author[0000-0003-0710-9474]{J.~L.~Marshall}
\affiliation{George P. and Cynthia Woods Mitchell Institute for Fundamental Physics and Astronomy, and Department of Physics and Astronomy, Texas A\&M University, College Station, TX 77843,  USA}

\author[0000-0002-6610-4836]{R.~Miquel}
\affiliation{Instituci\'o Catalana de Recerca i Estudis Avan\c{c}ats, E-08010 Barcelona, Spain}
\affiliation{Institut de F\'{\i}sica d'Altes Energies (IFAE), The Barcelona Institute of Science and Technology, Campus UAB, 08193 Bellaterra (Barcelona) Spain}

\author{J.~J.~Mohr}
\affiliation{Faculty of Physics, Ludwig-Maximilians-Universit\"at, Scheinerstr. 1, 81679 Munich, Germany}
\affiliation{Max Planck Institute for Extraterrestrial Physics, Giessenbachstrasse, 85748 Garching, Germany}

\author{R.~Morgan}
\affiliation{Physics Department, 2320 Chamberlin Hall, University of Wisconsin-Madison, 1150 University Avenue Madison, WI  53706-1390}

\author[0000-0003-2120-1154]{R.~L.~C.~Ogando}
\affiliation{Observat\'orio Nacional, Rua Gal. Jos\'e Cristino 77, Rio de Janeiro, RJ - 20921-400, Brazil}

\author{F.~Paz-Chinch\'on}
\affiliation{Center for Astrophysical Surveys, National Center for Supercomputing Applications, 1205 West Clark St., Urbana, IL 61801, USA}
\affiliation{Institute of Astronomy, University of Cambridge, Madingley Road, Cambridge CB3 0HA, UK}

\author[0000-0001-9186-6042]{A.~Pieres}
\affiliation{Laborat\'orio Interinstitucional de e-Astronomia - LIneA, Rua Gal. Jos\'e Cristino 77, Rio de Janeiro, RJ - 20921-400, Brazil}
\affiliation{Observat\'orio Nacional, Rua Gal. Jos\'e Cristino 77, Rio de Janeiro, RJ - 20921-400, Brazil}

\author{M.~Rodriguez-Monroy}
\affiliation{Centro de Investigaciones Energ\'eticas, Medioambientales y Tecnol\'ogicas (CIEMAT), Madrid, Spain}

\author[0000-0002-9646-8198]{E.~Sanchez}
\affiliation{Centro de Investigaciones Energ\'eticas, Medioambientales y Tecnol\'ogicas (CIEMAT), Madrid, Spain}

\author{V.~Scarpine}
\affiliation{Fermi National Accelerator Laboratory, P. O. Box 500, Batavia, IL 60510, USA}

\author{S.~Serrano}
\affiliation{Institut d'Estudis Espacials de Catalunya (IEEC), 08034 Barcelona, Spain}
\affiliation{Institute of Space Sciences (ICE, CSIC),  Campus UAB, Carrer de Can Magrans, s/n,  08193 Barcelona, Spain}

\author[0000-0002-1831-1953]{I.~Sevilla-Noarbe}
\affiliation{Centro de Investigaciones Energ\'eticas, Medioambientales y Tecnol\'ogicas (CIEMAT), Madrid, Spain}

\author[0000-0002-7047-9358]{E.~Suchyta}
\affiliation{Computer Science and Mathematics Division, Oak Ridge National Laboratory, Oak Ridge, TN 37831}

\author[0000-0003-1704-0781]{G.~Tarle}
\affiliation{Department of Physics, University of Michigan, Ann Arbor, MI 48109, USA}

\author[0000-0001-7836-2261]{C.~To}
\affiliation{Center for Cosmology and Astro-Particle Physics, The Ohio State University, Columbus, OH 43210, USA}

\author[0000-0001-7211-5729]{D.~L.~Tucker}
\affiliation{Fermi National Accelerator Laboratory, P. O. Box 500, Batavia, IL 60510, USA}

\author[0000-0002-8282-2010]{J.~Weller}
\affiliation{Max Planck Institute for Extraterrestrial Physics, Giessenbachstrasse, 85748 Garching, Germany}
\affiliation{Universit\"ats-Sternwarte, Fakult\"at f\"ur Physik, Ludwig-Maximilians Universit\"at M\"unchen, Scheinerstr. 1, 81679 M\"unchen, Germany}

\collaboration{69}{(DES Collaboration)}

\correspondingauthor{Sidney Mau, Ethan O.~Nadler}
\email{smau@stanford.edu, enadler@carnegiescience.edu}

\begin{abstract}
We use a recent census of the Milky Way (MW) satellite galaxy population to constrain the lifetime of particle dark matter (DM).
We consider two-body decaying dark matter (DDM) in which a heavy DM particle decays with lifetime $\tau$ comparable to the age of the universe to a lighter DM particle (with mass splitting $\epsilon$) and to a dark radiation species.
These decays impart a characteristic ``kick velocity,'' $V_{\mathrm{kick}}=\epsilon c$, on the DM daughter particles, significantly depleting the DM content of low-mass subhalos and making them more susceptible to tidal disruption.
We fit the suppression of the present-day DDM subhalo mass function (SHMF) as a function of $\tau$ and $V_{\mathrm{kick}}$ using a suite of high-resolution zoom-in simulations of MW-mass halos, and we validate this model on new DDM simulations of systems specifically chosen to resemble the MW.
We implement our DDM SHMF predictions in a forward model that incorporates inhomogeneities in the spatial distribution and detectability of MW satellites and uncertainties in the mapping between galaxies and DM halos, the properties of the MW system, and the disruption of subhalos by the MW disk using an empirical model for the galaxy--halo connection. By comparing to the observed MW satellite population, we conservatively exclude DDM models with $\tau < 18\Gyr$ ($29\Gyr$) for $V_{\mathrm{kick}}=20\kms$ ($40\kms$) at $95\%$ confidence.
These constraints are among the most stringent and robust small-scale structure limits on the DM particle lifetime and strongly disfavor DDM models that have been proposed to alleviate the Hubble and $S_8$ tensions.
\end{abstract}

\keywords{\href{http://astrothesaurus.org/uat/353}{Dark matter (353)}; \href{http://astrothesaurus.org/uat/1049}{Milky Way dark matter halo (1049)}; \href{http://astrothesaurus.org/uat/574}{Galaxy abundances (574)}}

\reportnum{DES-2021-0652}
\reportnum{FERMILAB-PUB-21-690-PPD}
\reportnum{SLAC-PUB-17647}


\section{Introduction}
\label{sec:intro}

The $\Lambda$ cold dark matter (CDM) paradigm---in which dark energy is a cosmological constant and dark matter (DM) is stable, cold, and collisionless---has been established as the concordance model of cosmology, accurately predicting the detailed content and structure of the universe throughout cosmic history \citep{DES:2018b,1807.06209}.
However, several recent cosmological tensions---namely the ``Hubble tension,'' concerning the present-day expansion rate of the universe, $H_0$, and the ``$\sigma_8$ tension,'' concerning the amplitude of matter clustering on quasi-linear scales---potentially point to new physics beyond $\Lambda$CDM (see \citealt{1907.10625} for a review). To address these tensions, it is essential to stress-test every assumption underlying the $\Lambda$CDM model, including the nature of the DM particle.

Any viable DM model must accurately predict both the expansion history of the universe and the formation of structure throughout cosmic history.
To date, all cosmological observations are consistent with a DM particle that is stable against decays.
However, there is a class of models that allow DM to decay, either to other dark sector particles or to Standard Model particles, which is consistent with current observational constraints.
These decays often transfer energy from DM to radiation-like species in the late-time universe, increasing the present-day expansion rate relative to its extrapolated evolution based on CMB measurements (e.g., \citealt{1903.06220,2006.03678}).
Intriguingly, recent studies of models where the DM particle decays with a lifetime comparable to the Hubble time suggest that DM decays can potentially alleviate the $H_0$ and/or $\sigma_8$ tensions (e.g., \citealt{1505.05511,1902.10636,1903.06220,2004.07709,2004.06114,2008.09615,Abellan210212498,2110.09562,2006.03678}).

In this paper, we focus specifically on DM models where late-time decays into other dark sector particles alter the velocity distribution of DM, which we refer to generically as decaying dark matter (DDM).
While these DDM models have predominantly been studied using expansion history and large-scale structure probes \citep[e.g., see][]{1201.2426,2011.04606}, cosmic structure on smaller scales offers a unique window into the microphysics of DM decays. In particular, because viable DDM lifetimes are comparable to the age of the universe, the effects of DM decays accumulate over time as matter perturbations grow in the post-recombination epoch. Late-time decays thus suppress the clustering of matter on small scales at late times; this effect has been leveraged to constrain DDM models using Lyman-$\alpha$ forest data \citep[e.g.,][]{0405013,1309.7354}.

DM decays impart momentum to the daughter particles, injecting kinetic energy into virialized DM halos. These momentum ``kicks''---which depend on the details of the decay mechanism---gradually reduce the central density of cuspy DM halos and deplete halos of mass (e.g., \citealt{1003.0419}). This process also makes subhalos more susceptible to tidal disruption as they orbit within a host halo, leading to the suppression of the abundance of low-mass halos and subhalos in DDM models at late times. This manifests as a deficit of small-scale structure, including faint satellite galaxies, relative to $\Lambda$CDM predictions \citep{1003.0419,1009.1912,Wang:2014} and also alleviates potential tensions between the predicted and inferred central densities of bright dwarf galaxies (the ``too big to fail'' problem; \citealt{Boylan-Kolchin11030007,Boylan-Kolchin11112048,Garrison-Kimmel14045313,Papastergis14074665}).

The population of Milky Way (MW) satellite galaxies, which contains the faintest observed galaxies in the universe, therefore offers a unique testing ground for DDM. Faint dwarf galaxies orbiting the MW occupy the smallest DM halos directly associated with galaxies and are the most dark matter-dominated systems known \citep[see a recent review by][]{1901.05465}. DM microphysics that impacts the formation, late-time abundance, and internal structure of small DM halos and subhalos can therefore be inferred from the abundance and properties of MW satellites.

Here, we derive robust and stringent constraints on the DM particle lifetime using a state-of-the-art census of the MW satellite population. Although DDM has previously been studied using MW satellites (e.g., \citealt{1009.1912,Wang:2014}), our work is the first to constrain it by leveraging observations over nearly the full sky, including the population of ultra-faint dwarf galaxies recently discovered by the Dark Energy Survey (DES; \citealt{astro-ph/0510346,1601.00329}).
Following the procedure developed in \cite{PaperI} and \cite{PaperII,PaperIII}, we combine (i) the observed population of MW satellites in DES and Pan-STARRS1 (PS1; \citealt{1612.05560}), (ii) observational selection functions derived from MW satellite searches in DES and PS1 data, (iii) a detailed forward model of the connection between MW satellite galaxies and DM subhalos, (iv) high-resolution cosmological simulations of MW-mass halos in DDM cosmologies, and (v) a Bayesian inference framework to constrain the DM particle lifetime. This method was used to constrain microphysical DM properties including its primordial velocity distribution, Standard Model coupling, and particle mass in \cite{PaperIII}.

Unlike previous DDM studies, our model only relies on the inferred abundance of low-mass subhalos above a minimum mass threshold, rather than their inferred central densities, which makes our constraints complementary to previous studies.
This minimum mass threshold corresponds to the peak mass of the smallest halo inferred to host MW satellites observed in the DES and PS1 data for CDM \citep{PaperII}, and therefore represents a conservative upper limit on this quantity for DDM, which further reduces halo masses as explored below.
In turn, our results provide a foundation for future work that combines satellite abundances with stellar velocity dispersion measurements to search for signatures of DM decays. Similar approaches will be useful to test other DM properties that simultaneously modify satellite abundances and density profiles at an observable level, including DM self-interactions \citep[e.g.,][]{Vogelsberger151205349,Tulin170502358, Salucci:2019,Drlica-Wagner:2019,2021arXiv210609050K}.

This paper is outlined as follows. In Section \ref{sec:ddm_models}, we describe our fiducial DDM model. We describe the impact of DDM physics and analytically estimate subhalo disruption timescales in Section \ref{sec:analytic_disruption}. In Section \ref{sec:shmf}, we derive a fitting function for the suppression of the subhalo mass function using cosmological zoom-in simulations of MW-like halos in DDM cosmologies. We incorporate this suppression in a forward model of the MW satellite population to derive DDM constraints in Section \ref{sec:constraints}. We compare our results to other small-scale and cosmological probes of DDM in Section \ref{sec:comparison}, and we conclude in Section \ref{sec:conclusion}. With the exception of Section \ref{sec:analytic_disruption}, we work in units with $c=1$. Furthermore, ``log'' refers to the base-$10$ logarithm.


\section{Decaying Dark Matter Overview}
\label{sec:ddm_models}

In this section, we outline classes of DM models that undergo decays, connect common model assumptions to cosmological observables, and describe the particular model of DDM that we consider in this work.
We emphasize that the DDM models we describe are phenomenological, rather than first-principles particle physics descriptions; models that feature scattering between DM and other particles are broadly categorized under self-interacting and/or interacting DM.

\subsection{Decaying Dark Matter Phenomenology}

DM models featuring decays typically involve massive parent DM particles, a fraction of which undergo particle decay to some number of daughter particles with a given decay lifetime.
Accordingly, decaying DM models can be broadly differentiated based on the following characteristics:
\begin{enumerate}
    \item The lifetime of the decay;
    \item The number of particles involved in each DM decay, and their masses; and
    \item Whether the decays exclusively involve dark sector particles or involve any Standard Model particles.
\end{enumerate}
DDM is typically regarded as a modification to the CDM paradigm rather than a completely distinct model; for instance, superWIMPs \citep{0302215,0306024,0403164} are an example of decaying CDM \citep{0005206}.
However, each of the assumptions listed are associated with specific cosmological signatures that distinguish them from stable CDM. 
We now consider the cosmological effects of each assumption in turn.


The lifetime of the DM decay sets the onset and rapidity of decays.
For extremely short lifetimes compared to the age of the universe, with $\tau\lesssim \mathcal{O}(\rm yr)$ \citep{Kaplinghat0507300}, decays before recombination can transition a fraction of the dark matter into a ``warm'' state by introducing an additional nonthermal velocity component; this can change the initial conditions for structure formation and increase the effective number of relativistic degrees of freedom in the early universe (e.g., \citealt{2004.06114} and references therein).
For decays that occur after recombination but on a shorter timescale than the age of the universe, cosmological observables are generally affected by the resulting changes in DM composition at late times.
For extremely long decay lifetimes---of the order the age of the universe or longer---the observable impacts relative to a stable DM model are minimal.
Furthermore, note that the lifetime of the decay is occasionally parameterized according to the fraction of DM that has undergone decays by $z=0$ \citep[e.g.,][]{0005206,0305496}.

The number of particles involved in the decay has several consequences for cosmological observables.
By allowing for the parent particle to decay into $N$ products, it is possible to introduce up to $N$ new and potentially unique dark sector particles.
The case of $N = 2$ allows for the decay products to take on a range of mass splitting values; our fiducial model (detailed below) is an example of such a decay in which one of the decay products is a dark radiation component, and the other is a massive particle of comparable mass to the parent particle.
For $N > 2$, the complexity of the model increases (e.g., as in \citealt{1410.0683,1510.06026,2004.07709}), leading to cosmological signatures that are generally more difficult to disentangle than the $N=2$ case.

Finally, it is interesting to consider whether or not the decay products include Standard Model particles.
Particle collider experiments have been used to constrain dark matter decays that operate via Standard Model portals and scattering mechanisms, including the Higgs portal \citep[e.g.,][]{2011.05259} or a more general coupling in the case of pseudo-Dirac dark mater \citep[e.g.,][]{1806.05185,1812.03169,2108.13422}.
Similarly, gamma-ray and X-ray telescopes constrain WIMP-like models that decay into Standard Model particles \citep[e.g.,][]{1506.00013,Gaskins160400014,2103.13242}.

\subsection{Fiducial Decaying Dark Matter Model}

In this paper, our fiducial DDM model consists of a cold, massive DM particle $\chi$ that decays to a slightly less massive daughter DM particle $\chi'$ and a massless dark radiation species $\gamma'$,
\begin{equation}
    \chi \rightarrow \chi' + \gamma'.
    \label{eq:decay_process}
\end{equation}
In our fiducial cosmology, $\chi$ composes the entirety of the initial DM.
Denoting the mass of $\chi$ by $M$ and the mass of $\chi'$ by $m$, this model can be parameterized by the decay lifetime $\tau$ and the mass splitting \citep{Wang:2014}
\begin{equation}
    \epsilon \equiv \frac{M-m}{M}.
\end{equation}
These decays impart a recoil velocity
\begin{equation}
    V_{\mathrm{kick}} \approx  \epsilon,
\end{equation}
assuming $V_{\rm kick} \ll 1$, on the daughter DM particle in the center-of-mass rest frame of the parent particle (recall that we work in natural units with $c=1$). Following \cite{1001.3870} and \cite{Wang:2014}, we consider a parameter space in which the decays occur at late times with small recoil velocities, with $\tau\sim \mathcal{O}(\Gyr)$ and $\epsilon\ll 1$. Such models only mildly affect large-scale structure (e.g., \citealt{Poulin160602073}) while yielding potentially observable effects on smaller scales, and particularly on low-mass DM halos.
Note that our fiducial DDM model makes no assumptions about interactions between the dark sector (i.e., any of the particles $\chi$, $\chi'$, or $\gamma'$) and the Standard Model.

In Section \ref{sec:shmf-fit}, we demonstrate that the late-time suppression of DDM subhalo abundances can be fit reasonably well using a functional form similar to that used to describe the suppression of the subhalo mass function in warm dark matter (WDM). However, we emphasize that late-time subhalo disruption is a dynamical effect in our DDM models while, for WDM-like models, the effect is imprinted on the linear matter power spectrum at early times.
This distinction between DDM and WDM models can also lead to differences in the relative abundance of low-mass isolated halos and subhalos at a given mass scale; in particular, the abundance of isolated halos and subhalos is roughly equally suppressed in WDM-like models \citep{Stucker210909760}, while subhalo abundances can be preferentially suppressed in DDM.

It is informative to consider the limiting cases of our fiducial two-body decay model.
Longer DDM lifetimes lead to fewer DM particle decays, and the model becomes more similar to CDM at arbitrarily later times. Similarly, a smaller mass splitting between parent and daughter particles leads to smaller kick velocities---and less energy carried away by the dark radiation component---and the model again approaches CDM.
We will consider lifetimes comparable to the Hubble time (or longer) and kick velocities comparable to the circular velocities of subhalos that host MW satellite galaxies; thus, our analysis only directly constrains decays that change the distribution of DM structure relative to CDM at late times.


\section{Effects of  Dark Matter Decays on Milky Way Satellites}
\label{sec:analytic_disruption}

To develop intuition for the effects of dark matter decays on MW satellite galaxy abundances, we consider a simple toy model wherein we compute the maximum initial halo mass that remains above a $z=0$ mass threshold corresponding to the subhalos that host the faintest galaxies we use in our likelihood analysis.
This calculation provides a rough estimate of the halo mass scales that are significantly affected in the DDM models we consider; however, we use the cosmological zoom-in simulations presented in Section \ref{sec:shmf} to make precise predictions for DDM subhalo abundances.

We note that, although the toy model presented in this section applies equally to subhalos and isolated halos, subhalos experience tidal effects (after infall) that accelerate their mass loss and that are exacerbated for DDM subhalos with reduced central densities (\citealt{1001.3870}; also see Figure \ref{fig:vmax}).
Thus, our estimates of the halo masses that are significantly affected by DM decays represent conservative lower limits for the subhalo masses that are impacted, and the combination of decays and tidal stripping shapes the simulated DDM subhalo populations we study in Section \ref{sec:shmf}.
We reinstate units of $c$ in this section for clarity.

For a halo with an initial mass $M_i$ at time $t_i$, it is possible to determine the time needed for decays to reduce the halo mass below a given threshold.
By choosing this threshold to correspond to the minimum mass of subhalos that host observed MW satellites, this calculation yields a rough estimate of the impact of DM decays on MW satellite abundances.
In addition, it highlights an important difference between DDM and other non-CDM models, including WDM, which are commonly considered on dwarf galaxy scales. In DDM, low-mass halos can form at early times but evaporate due to decays by late times, whereas, in models that only suppress the linear matter power spectrum (e.g., WDM), subhalos below a mass threshold never form.

For a halo with an initial number of particles $N_i$ that decay with a lifetime $\tau$, the number of particles that have \emph{not} undergone decay within a time interval $\Delta t \equiv t_f - t_i$ is given by
\begin{equation}
    N(t) = \exp\left[-\frac{\Delta t}{\tau}\right] N_i,
\end{equation}
assuming that all daughter particles are ejected from the halo due to the recoil kick velocity.
Because we will analyze DDM models with $V_{\rm kick}=20\kms$ and $40\kms$, this is a reasonable assumption for the subhalos that drive our constraints, which typically have peak virial masses of ~$\sim 10^8\Msun$ and peak maximum circular velocities of $V_{\rm peak}\approx 20\kms$ \citep{Nadler180905542,PaperII}, corresponding to escape velocities of $V_{\rm esc}\approx28\kms$. In the regime of $V_{\rm kick} \gtrsim V_{\rm peak}$ (which guarantees that $V_{\rm kick} \gtrsim V_{\rm max}$), the impact of decays on low-mass halos is effectively determined by the DM lifetime alone \citep{Wang:2014}.\footnote{This is consistent with a calculation of the energy injection due to DM decays following \cite{0806.0602}, which indicates that decays are unlikely to unbind halos of the masses we consider within a Hubble time.}

\begin{figure}
\includegraphics[width=\columnwidth]{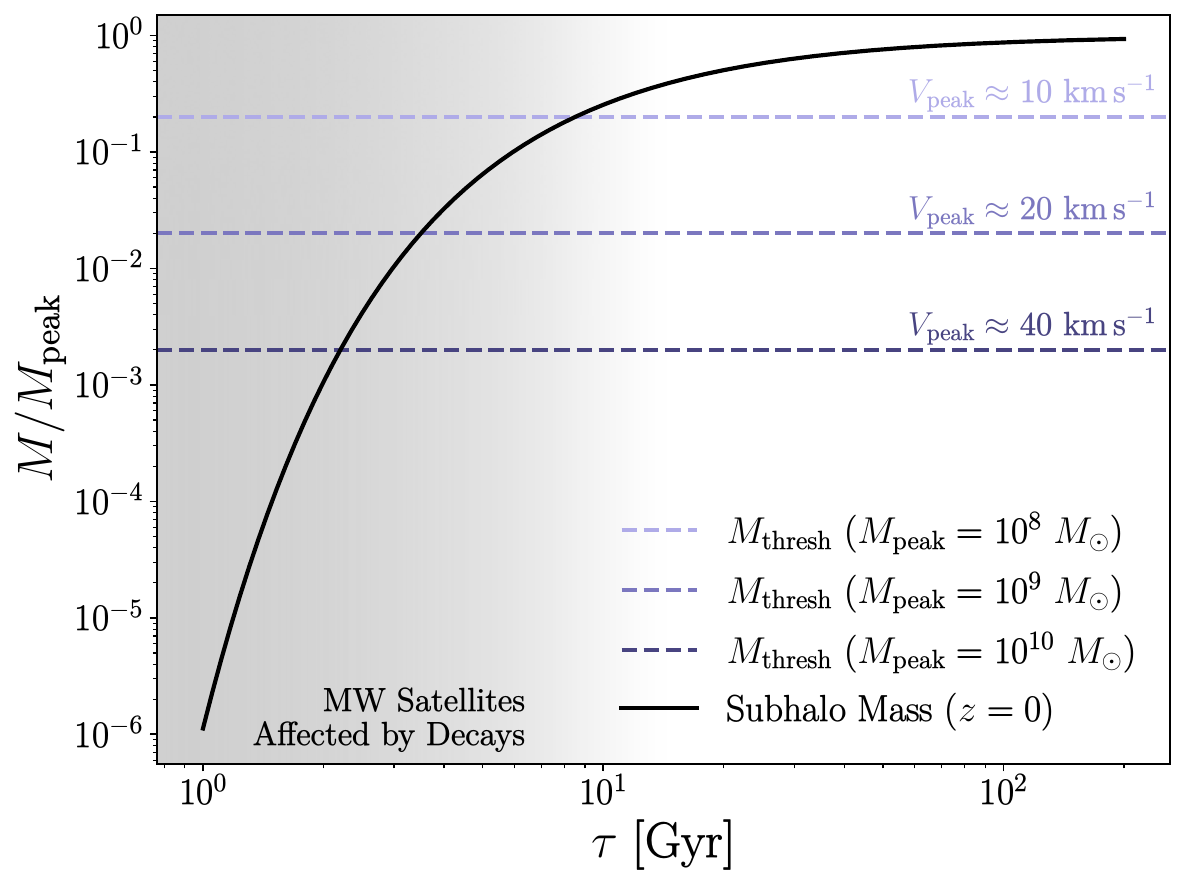}
    \caption{
        Final subhalo mass (in units of peak virial mass) vs.\ dark matter decay lifetime for a subhalo that undergoes decays for the age of the universe ($t_H = 13.7 \Gyr$). For each peak subhalo mass, dashed horizontal lines indicate a mass threshold $M_{\rm thresh} = 2 \times 10^7 \Msun$ that we assume corresponds to the minimum mass for subhalos that host satellites galaxies. This mass threshold corresponds to subhalos resolved with roughly $100$ particles in our zoom-in simulations (see Section \ref{sec:shmf-sims}). For decay lifetimes of $\lesssim 10 \Gyr$, the abundance of low-mass ($M_{\rm peak}\sim 10^8\Msun$) subhalos above the mass threshold is significantly affected by decays alone (gray region).
    }
\label{fig:disrupt}
\end{figure}

We also assume that all parent and all daughter DM particles, respectively, have the same mass; hence, the total mass of the halo after time interval $\Delta t$ is then given by
\begin{equation}
    M_{\rm sub}(\Delta t) = \exp\left[-\frac{\Delta t}{\tau}\right] M_{i}.
\end{equation}
To facilitate comparison with our simulation results, we adopt the peak virial mass $M_{\rm peak}$, defined as the largest virial mass a subhalo attains along its main branch in our zoom-in simulations (see Section \ref{sec:shmf-sims}), as the initial mass $M_i$.
For simplicity and because $M_{\rm peak}$ is achieved at fairly high redshift for typical subhalos in our simulations, we adopt the age of the universe $t_H$ as $\Delta t$.
Neither assumption has an important effect on the qualitative results of our mass-loss calculation, which is shown in the left panel of \figref{disrupt}.

The majority of observed MW satellites that enter our analysis inhabit subhalos in the peak virial mass range of $10^8 \Msun \lesssim M_\mathrm{peak}\lesssim 10^{10} \Msun$, where $M_\mathrm{peak}$ is the largest virial mass each subhalo ever achieves over its entire main-branch history \citep{PaperII}.
In this calculation and in our comparison to the MW satellite population, we additionally assume that halos that fall below a minimum mass threshold at $z=0$ cannot host observed MW satellites.
In our cosmological DDM simulations, subhalos are no longer identified when their mass falls below a resolution limit of $\approx 2\times 10^7\Msun$, corresponding to roughly $100$ particles (see Section \ref{sec:shmf-sims}); we adopt this as our minimum subhalo mass threshold for MW satellites, $M_{\mathrm{thresh}}$. This mass threshold is reasonable given the dynamical properties of observed MW satellites (e.g., \citealt{Strigari08083772}).

Imposing our minimum $z=0$ mass threshold yields a condition for the maximum $M_{\rm peak}$ that is significantly impacted by decays:
\begin{equation}
    M_{\rm sub}(t_H) \leq M_{\rm thresh}\implies
    \exp\left[-\frac{t}{\tau}\right] M_{\rm peak} \leq M_{\rm thresh}.
\end{equation}
\figref{disrupt} shows the relation between final halo mass (in units of $M_{\rm peak}$) and decay lifetime, while indicating our mass threshold over the range of peak subhalo masses relevant for our analysis. Thus, for decay lifetimes $\tau\lesssim 10\Gyr$, we expect low-mass subhalos ($M_{\rm peak}\sim 10^8\Msun$) to approach the threshold mass and (by construction) the resolution limit of our simulations due to mass loss from decays \emph{alone}. This foreshadows the suppression of subhalo abundances at these mass scales derived from our simulations in Section \ref{sec:ddm-sims}.
Again, we note that our mass-loss calculation applies equally to isolated halos, and thus predicts that the abundance of isolated halos with similar peak masses that remain above our threshold mass at $z=0$ is suppressed by decays. Appendix \ref{sec:appendix_c} examines this effect in our cosmological zoom-in simulations.

We reiterate that the results in \figref{disrupt} represent order-of-magnitude estimates of DDM effects on low-mass subhalos, and that our constraints are instead based on detailed cosmological DDM simulations (see Section \ref{sec:shmf}).
Nonetheless, the toy model presented above demonstrates that the abundance and structure of subhalos expected to host MW satellite galaxies are sensitive to DM particle lifetimes of $\mathcal{O}(10\Gyr)$ and to the microphysics of these decays encapsulated by $V_{\mathrm{kick}}$. 


\section{Decaying Dark Matter Subhalo Populations}
\label{sec:shmf}

The DDM models we consider differ from many popular alternatives to CDM in that they feature late-time suppression of small subhalos, rather than a suppression in the linear matter power spectrum, which eventually manifests as a reduction of subhalo abundances (i.e., smaller subhalos simply do not form in, for example, WDM, whereas in DDM they can form initially but are disrupted by late times).
Importantly, this precludes a first-principles mapping between the impact of DDM and other well-studied CDM alternatives like WDM on small-scale structure formation.
Instead, we study the effects of DM decays at late times and their impact on subhalo populations using cosmological zoom-in simulations of MW-mass systems to model the nonlinear impact of late-time DM decays. We use these simulations to fit the suppression of the subhalo mass function as a function of $\tau$ for $V_{\mathrm{kick}}=20\kms$ and $40\kms$ separately, and we use these fits in our likelihood framework to constrain $\tau$ as described in Section \ref{sec:results}.\footnote{We focus on $V_{\mathrm{kick}}=20\kms$ and $40\kms$ following \cite{1001.3870} and \cite{Wang:2014}, who found that these kick velocities are sufficient to significantly impact subhalo abundances and internal densities.}

Section \ref{sec:shmf-sims} gives an overview of the DDM and CDM simulations used in this work, Section \ref{sec:ddm-sims} describes the general characteristics of the DDM subhalo populations, and Section \ref{sec:shmf-fit} studies the suppression of subhalo abundances relative to CDM and derives an analytic model for this effect to enable the statistical inference performed in \secref{constraints}.

\subsection{Cosmological Zoom-in Simulations}
\label{sec:shmf-sims}

To study the impact of DDM physics on low-mass subhalos and to derive predictions for the suppression of the subhalo mass function (SHMF), we use an expanded set of cosmological zoom-in simulations of MW-mass halos in DDM models based on those presented in \citet{Wang:2014}. In particular, we study six simulations with DDM parameters $(\tau/\Gyr, V_{\rm kick}/\kms) \in \{(10, 20),\allowbreak (20, 20),\allowbreak (20, 40),\allowbreak (40, 20),\allowbreak (40, 40),\allowbreak (80, 40)\}$ and a corresponding CDM simulation, all with identical initial conditions.
These simulations were performed using a version of the \code{GADGET-2} \citep{Springel:2005} and \code{GADGET-3} N-body codes as modified by \citet{1003.0419}. The simulations are run in a box of length $50\h\Mpc^{-1}$ per side and zoom in on a halo mass of $M \approx 10^{12}\Msun$ \citep{Wang:2014}. The highest-resolution region is simulated with a Plummer-equivalent force softening of $143\pc$ and a particle mass of $1.92\times 10^5\Msun$.\footnote{We define virial quantities according to the \cite{Bryan_1998} virial definition, with overdensities $\Delta_{\rm vir}$ set according to the cosmological parameters in our two simulation suites.} Cosmological parameters are set to $\Omega_M=0.266$, $\Omega_\Lambda=0.734$, $n_s=0.963$, $h=0.71$, and $\sigma_8=0.801$ \citep{WMAP7}, and we analyze these simulations using the \code{ROCKSTAR} \citep{1110.4372} halo finder and \code{CONSISTENT-TREES} \citep{1110.4370} merger tree code.

We emphasize that the \cite{Wang:2014} simulations are not specifically selected to resemble the MW in properties beyond its host halo mass. To derive the DDM constraints in \secref{constraints}, we therefore analyze the two additional MW-like N-body zoom-in simulations originally presented in \cite{Mao:2015} and studied in \cite{PaperII,PaperIII} to model the MW satellite galaxy population. The properties of these host halos are consistent with the mass and concentration of the MW halo and include both realistic Large Magellanic Cloud (LMC) analog systems on recent infall and \textit{Gaia}-Sausage-Enceladus-like merger events at early times \citep{Belokurov:2018,Helmi:2018}. We describe these simulations in detail in Appendix \ref{sec:appendix_a}. We also perform several DDM resimulations of these systems, which we use in Appendix \ref{sec:appendix_b} to validate that the impact of DDM on the subhalo populations in our expanded suite of \cite{Wang:2014} simulations is consistent with its impact on MW-like systems.
Throughout this paper, the expanded suite of simulations from \cite{Wang:2014} are represented with a yellow--green--blue--purple color scheme.
Plots using the MW-like resimulations only appear in the appendix and are represented with a red--purple color scheme.

\subsection{Decaying Dark Matter Subhalo Populations}
\label{sec:ddm-sims}

\begin{figure*}
\includegraphics[width=\textwidth]{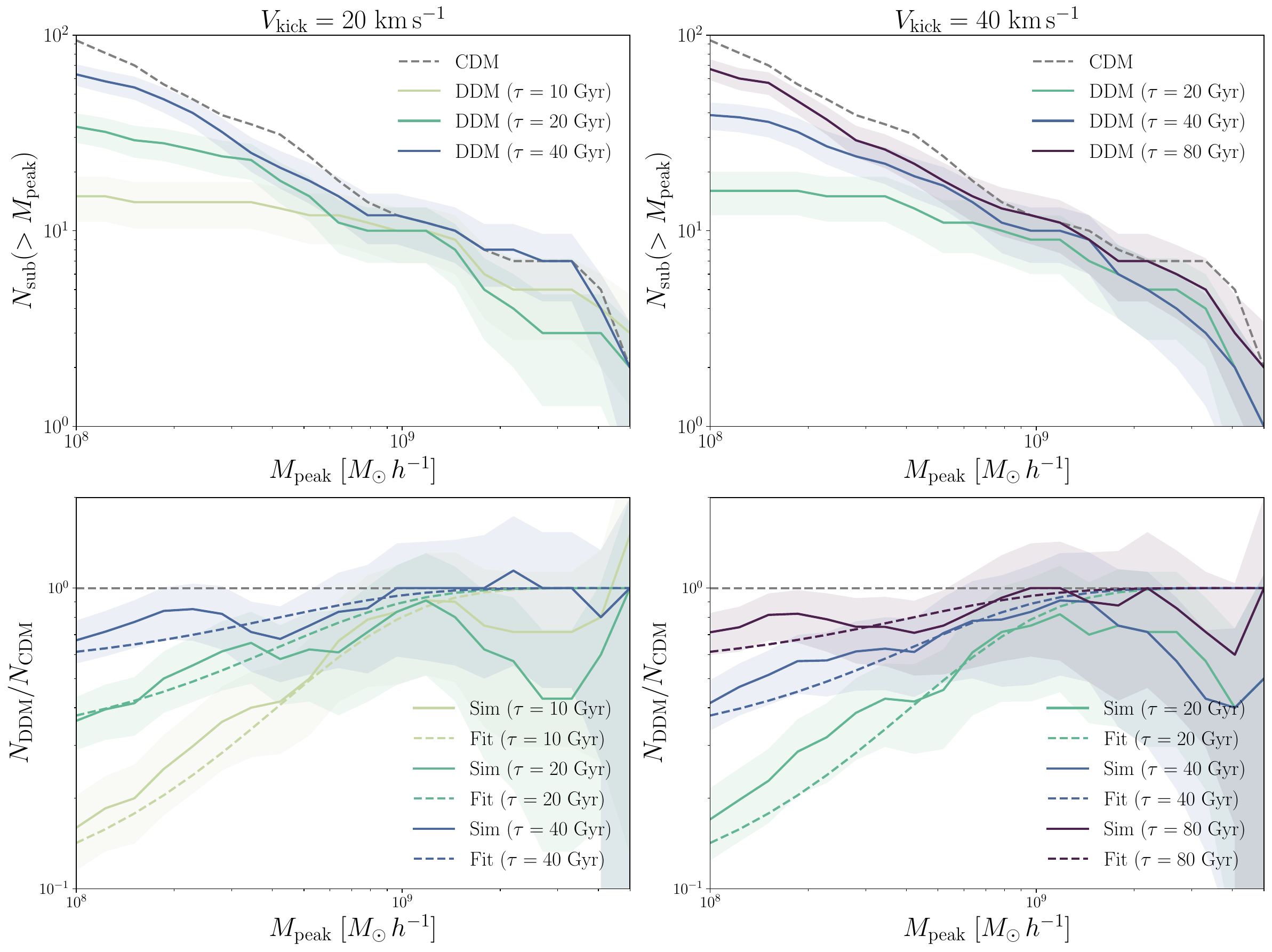}
\caption{\emph{Top panels}: subhalo mass functions at $z=0$ for the MW-mass host halo in the expanded suite of zoom-in simulations from \cite{Wang:2014}. SHMFs are shown at $z=0$ in CDM (dashed gray) and in DDM models with $\tau = 10$, $20$, $40$, and $80\Gyr$ (from yellow to purple), with $V_{\mathrm{kick}}=20\kms$ (left panel) and $40\kms$ (right panel) as a function of peak subhalo virial mass $M_{\mathrm{peak}}$. All SHMFs are restricted to subhalos above a conservative resolution threshold of $V_{\rm peak} > 10 \kms$ and $V_{\rm max} > 9 \kms$. Shorter DDM decay lifetimes result in fewer surviving subhalos at low peak masses relative to CDM, and this effect is more pronounced for models with higher kick velocities. \emph{Bottom panels}: suppression of the subhalo mass function relative to CDM at $z=0$ for the same DDM models shown in the top panels. Solid lines show the SHMF suppression measured directly from our expanded suite of zoom-in simulations based on \cite{Wang:2014}, and dashed lines show the best-fit function derived in Section \ref{sec:shmf-fit} for each DDM model. Shaded bands indicate $68\%$ confidence interval Poisson uncertainties on the simulation measurements, which are consistent with no suppression (i.e., $N_{\mathrm{DDM}}/N_{\mathrm{CDM}} = 1$ at peak masses $M_{\mathrm{peak}}\gtrsim 10^9\Msun$. Thus, our fitting functions approach unity in the high-mass regime.)}
\label{fig:shmf}
\end{figure*}

We begin by summarizing the main differences between the DDM and CDM subhalo populations in the \cite{Wang:2014} simulations. \figref{shmf} shows the SHMFs in the \cite{Wang:2014} simulations introduced in \secref{shmf-sims}, defined as the cumulative abundance of subhalos as a function of peak subhalo virial mass.
The SHMFs for DDM models with a range of lifetimes for each $V_{\rm kick} = 20\kms$ ($40\kms$) are shown on the upper left (right) panels.
Lowering the decay lifetime and increasing the kick velocity both systematically decrease the abundance of surviving subhalos at low masses. Note that the ratio of the DDM SHMF relative to CDM at peak subhalo masses above $\sim 10^9\Msun$ is consistent with unity within the statistical precision of our simulations; this behavior is conservatively reflected in our fitting function predictions (lower panels of \figref{shmf}; see \secref{shmf-fit}). We leave a detailed study of dark matter decays on the abundance and density profiles of more massive subhalos to future work. 
Importantly, our main results are driven by the abundance of subhalos with masses below this threshold (see \citealt{PaperII,PaperIII}).

\begin{figure*}
\includegraphics[width=\textwidth]{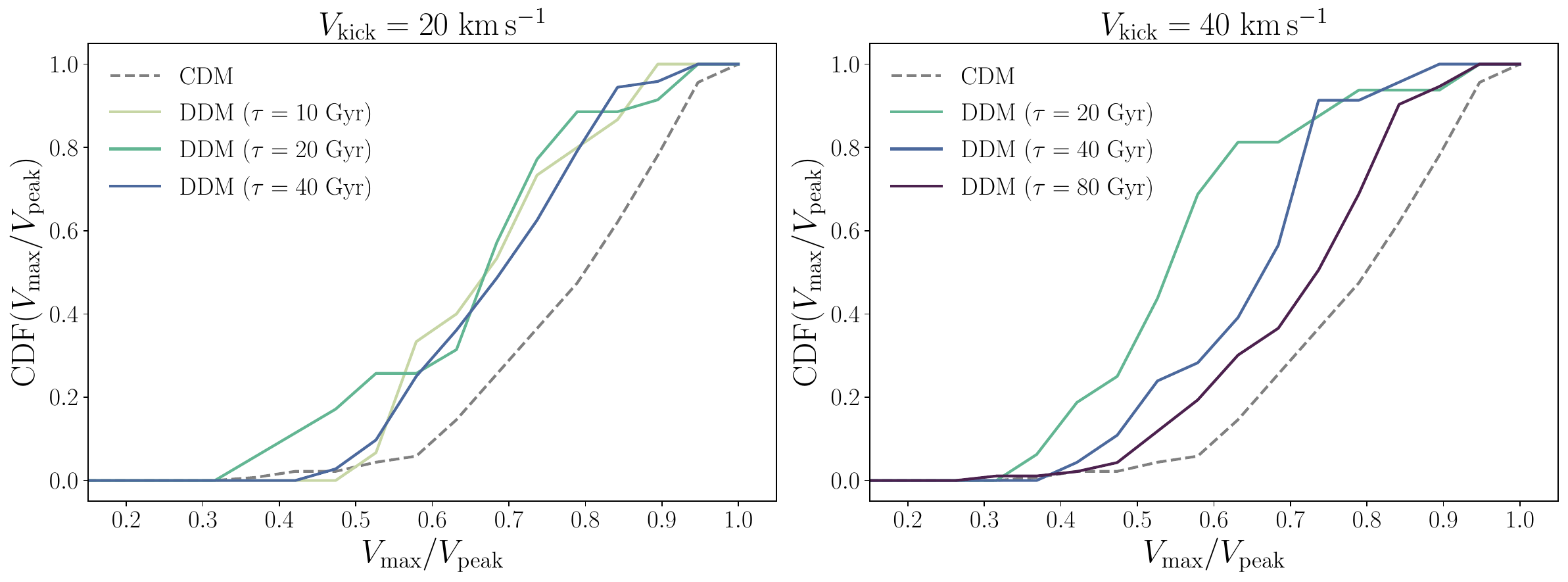}
\caption{Cumulative distribution function of subhalo $z=0$ maximum circular velocity, $V_{\mathrm{max}}$, relative to its peak value, $V_{\rm peak}$, for subhalos of the MW-mass host halo in the expanded suite of zoom-in simulations from \cite{Wang:2014}. Distributions are shown for CDM (gray) and in DDM models with $\tau = 10$, $20$, $40$, and $80\Gyr$ (from yellow to purple), with $V_{\mathrm{kick}}=20\kms$ (left column) and $40\kms$ (right column). All distributions are restricted to subhalos above a conservative resolution threshold of $V_{\rm peak} > 10 \kms$ and $V_{\rm max} > 9 \kms$. Subhalos in our DDM simulations exhibit systematically lower values of $V_{\mathrm{max}}$/$V_{\rm peak}$ relative to CDM, and this effect is more pronounced for shorter decay lifetimes and higher kick velocities. This indicates that the central densities of DDM subhalos are reduced, consistent with previous studies (e.g., \citealt{1003.0419,Wang:2014}).}
\label{fig:vmax}
\end{figure*}

\figref{vmax} shows the distribution of subhalo present-day maximum circular velocity, $V_{\rm max}$, divided by its peak value, $V_{\rm peak}$, again for DDM models with $V_{\rm kick} = 20\kms$ ($40\kms$) on the left (right).
There is a pronounced decrement in this ratio for each of our DDM simulations relative to their CDM counterparts, implying that subhalos' central densities are reduced in DDM, consistent with the findings of \cite{1003.0419} and \cite{Wang:2014}.
This effect could be leveraged to improve DDM constraints by incorporating MW satellites' stellar velocity dispersion measurements in the inference \citep[e.g.,][]{Drlica-Wagner:2019,2021arXiv210609050K}, rather than relying solely on their surface brightness function and radial distribution.

\begin{figure*}
\includegraphics[width=\textwidth]{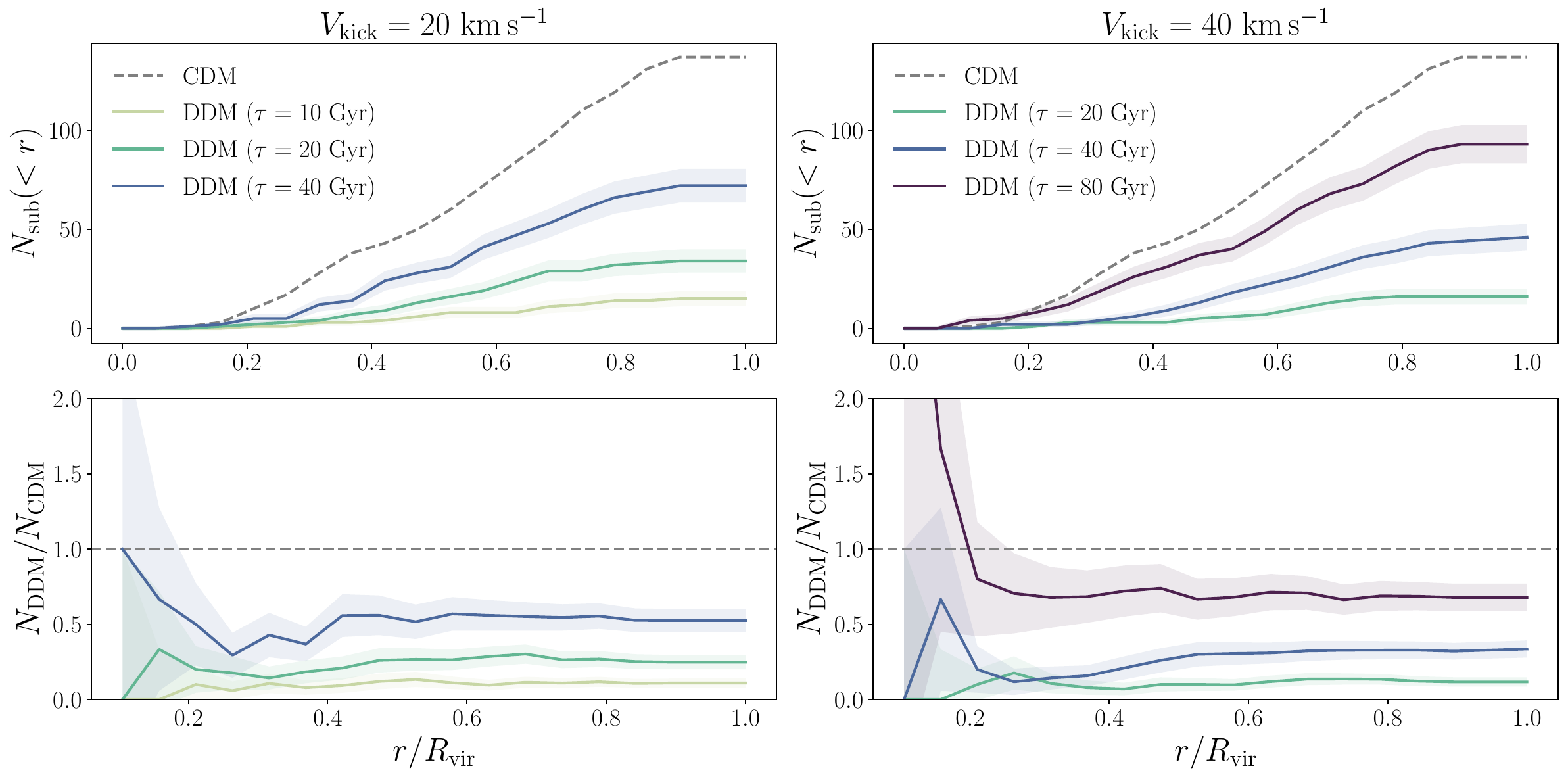}
\caption{Subhalo radial distributions at $z=0$ as a function of distance from the center of the host halo in the expanded suite of zoom-in simulations from \cite{Wang:2014} for CDM (dashed gray) and for DDM models with $\tau = 10$, $20$, $40$, and $80\Gyr$ (from yellow to purple), with $V_{\mathrm{kick}}=20\kms$ (left panel) and $40\kms$ (right panel). The bottom panels show the ratio of the radial distribution in each DDM model relative to CDM. All results are restricted to subhalos above a conservative resolution threshold of $V_{\rm peak} > 10 \kms$ and $V_{\rm max} > 9 \kms$. Shorter DDM decay lifetimes result in fewer surviving subhalos at all radii relative to CDM, and this effect is more pronounced for models with higher kick velocities. The virial radius is taken to be 300\kpc. Shaded bands indicate $68\%$ confidence interval Poisson uncertainties on the simulation measurements.}
\label{fig:shrf}
\end{figure*}

\figref{shrf} shows the $z=0$ radial distribution of surviving subhalos in the DDM and CDM simulations. Note that the suppression of subhalo abundances in these DDM simulations is roughly radially independent, which is consistent with the intuition developed in Section \ref{sec:analytic_disruption}: because decays alone significantly deplete the mass contained in small subhalos, the suppression of subhalo abundance is mainly determined by $M_{\mathrm{peak}}$, which is not strongly correlated with present-day radial distance (e.g., \citealt{Springel08090898}), even in the presence of the LMC \citep{Nadler210912120}.\footnote{This is also supported by Figures \ref{fig:redshift}--\ref{fig:redshift-isolated}, which show that the suppression of isolated halo and subhalo abundances in our DDM resimulations of MW-like systems is comparable.} On the other hand, note that suppression of subhalo abundances due to the Galactic disk is a strong function of radius and only weakly depends on subhalo mass \citep{Garrison-Kimmel170103792,Kelley181112413}.
We will exploit the fact that the shape of the radial distribution of surviving subhalos is approximately unchanged in DDM relative to CDM when deriving our constraints in \secref{constraints}.

In Appendix \ref{sec:appendix_c}, we study the evolution of DDM subhalo abundances in our MW-like resimulations described in Appendix \ref{sec:appendix_a}. We find that the suppression of the DDM SHMF relative to CDM sets in at late times ($z\lesssim 2$), which is expected based on the long timescale (of the order of the Hubble time) of the decays in the models we consider.
This is consistent with the intuition that small-scale structure is only suppressed in DDM at late times.
This implies that probes of small-scale structure in the low-redshift universe, including MW satellite galaxies, stellar streams, and strong gravitational lenses, are uniquely suited to constrain DDM models. On the other hand, probes of the same comoving scales at earlier times, including the Lyman-$\alpha$ forest \citep{Viel13062314,Irsic170304683,Palanque-Delabrouille191109073} and the high-redshift galaxy luminosity function \citep{Schultz14013769,Menci160101820,Corasaniti161105892,Rudakovskyi210404481}, offer relatively less constraining power for these models.

\subsection{Subhalo Mass Function Suppression}
\label{sec:shmf-fit}

Having explored the main features of the subhalo populations in our DDM simulations, we now derive a fitting function for the suppression of the SHMF relative to CDM. In particular, we express the DDM SHMF as
\begin{equation}
    \left(\frac{\mathrm{d}N_{\rm sub}}{\mathrm{d}M}\right)_{\rm DDM} \equiv f_{\rm DDM}(M,\tau,V_{\rm kick})\left(\frac{\mathrm{d}N_{\rm sub}}{\mathrm{d}M}\right)_{\rm CDM},\label{eq:shmf}
\end{equation}
where $f_{\rm DDM}(M,\tau,V_{\rm kick})$ is the suppression of the DDM SHMF relative to that in CDM as a function of subhalo peak virial mass, decay lifetime, and recoil kick velocity.
Note that, due to the resolution limit of our simulations, $f_{\mathrm{DDM}}$ describes the suppression of subhalo abundances above a present-day mass threshold of $\sim 2\times 10^7\Msun$.
To measure this quantity, we fit the output of the DDM simulations from \cite{Wang:2014} described above using a functional form of the SHMF similar to that proposed for warm dark matter \citep{2001.11013,2003.01125}:
\begin{equation}
    f_{\rm DDM}(M,\tau,V_{\rm kick}) = \left(1+\left(\frac{ M_0}{M}\right)^{\beta}\right)^{-\gamma(\tau,V_{\rm kick})},\label{eq:f_ddm}
\end{equation}
where 
\begin{equation}
\gamma(\tau,V_{\rm kick}) \equiv \gamma_0 \frac{V_{\rm kick}}{\tau}.
\end{equation}

To determine the parameters $M_0$, $\beta$, and $\gamma_0$, we perform least-squares fitting over a grid of SHMF values as a function of $M_{\rm peak}$ with varying $\tau$ and $V_{\rm kick}$.
We perform this fit over a mass range of $3.8 \times 10^7$--$9.4 \times 10^9 \Msun$ and derive best-fit values of
\begin{align}
    \log (M_0/\Msun) &= 8.5, \\ 
    \beta  &= 0.6, \\ 
    \gamma_0 &= 0.8, 
\end{align}
with a reduced $\chi^2$ goodness of fit of 0.3 (which reflects the relatively large Poisson uncertainties on the SHMFs).\footnote{The suppression of the DDM SHMF has a slightly weaker dependence on subhalo mass than in WDM, for which $\beta\approx 1.0$ and $\gamma \approx -0.99$ \citep{Lovell13081399}.} 
The fit of \eqnref{f_ddm} with these best-fit values is shown in the lower panels of \figref{shmf}. Although this fit slightly overpredicts SHMF suppression at peak masses below $\sim 5\times 10^8\Msun$, we estimate that this only influences our limits at the $\sim 5\Gyr$ level because most of the constraining power results from satellites associated with higher-mass halos (Section \ref{sec:results}).

It is important to note that, while \eqnref{f_ddm} empirically describes the suppression of the DDM SHMF relative to CDM, it is neither derived from first principles nor physically motivated. Moreover, it only accurately describes SHMF suppression for the range of $(\tau, V_{\rm kick})$ sampled in our expanded suite of simulations based on \cite{Wang:2014}, which is bounded below (above) by $\tau/\Gyr,V_{\rm kick}/\kms=10,20$ ($80,40$), with $z = 0$.


\section{Constraints from Milky Way Satellite Galaxies}
\label{sec:constraints}

\subsection{Forward Model and Fitting Procedure}
\label{sec:fitting}

To constrain the effects of DDM physics using observations of MW satellite galaxies, we incorporate the suppression of the SHMF derived from our DDM simulations into a forward-modeling framework that generates realizations of the satellite populations observed by DES and PS1. In particular, we apply the galaxy--halo connection model presented in \citet{PaperII} to the subhalo populations in the two CDM MW-like simulations from the \cite{Mao:2015} suite in order to predict the absolute magnitude, half-light radius, and Galactocentric distance distributions of satellites corresponding to subhalos in each simulation. We model the suppression of the SHMF in DDM by applying a fitting function to these CDM simulations because they have realistic LMC analogs and to enable a more direct comparison with \cite{PaperIII}.

Our galaxy--halo connection model includes eight free parameters that control the abundance-matching model that relates satellite luminosity to subhalo peak maximum circular velocity, the size model that relates satellite half-light radius to subhalo size at accretion, the efficiency of subhalo disruption due to the Galactic disk, and the minimum peak halo mass and scatter of the galaxy occupation fraction.
These parameters are defined in Appendix \ref{sec:appendix_d} and \tabref{model}; we refer the reader to \cite{PaperII} for a comprehensive description of the galaxy--halo connection model. Following \cite{PaperIII}, we add one free parameter to this model, $\tau$, which controls the suppression of the DDM SHMF at a given $V_{\mathrm{kick}}$ according to \eqnref{f_ddm}, and we perform the analysis for $V_{\mathrm{kick}}=20\kms$ and $40\kms$ separately. The contribution of each satellite to the mock observed number count is then weighted according to the probability that its corresponding subhalo survives for a given set of DDM parameters.

As in \cite{PaperIII}, this procedure assumes that the shape of the subhalo radial distribution is unchanged in DDM relative to CDM, which is demonstrated in \figref{shrf} for our MW-mass simulations and \figref{shrf-resim} for the MW-like simulations used in the inference. Furthermore, we do not modify the fiducial subhalo disruption probabilities predicted by the \cite{Nadler171204467} algorithm, which was calibrated on CDM hydrodynamic simulations. This is a conservative assumption because DDM reduces the central densities of surviving subhalos (see \figref{vmax}), making them more susceptible to tidal disruption; however, because these disruption probabilities are marginalized over in our fitting procedure, this assumption is not expected to significantly impact our constraints. Note that we do not account for adiabatic expansion of satellite sizes due to decays, which is also a conservative strategy because this effect would push some predicted satellites below the detectability threshold, thereby forcing even lower-mass subhalos to contribute and leading to more stringent DDM constraints.

For a given set of galaxy--halo connection and DDM model parameters, we perform mock observations of the DES and PS1 satellite populations using the observational selection functions presented in \citet{PaperI} by self-consistently orienting the survey footprints relative to the LMC analogs in our MW-like simulations. Thus, our procedure explicitly incorporates inhomogeneities in the spatial distribution and detectability of MW satellites and yields realizations of the observed DES and PS1 satellite populations that are compared to the data from \cite{PaperI} by assuming that satellite surface brightness is distributed according to a Poisson point process in each survey footprint.

\begin{figure*}
\includegraphics[width=\textwidth]{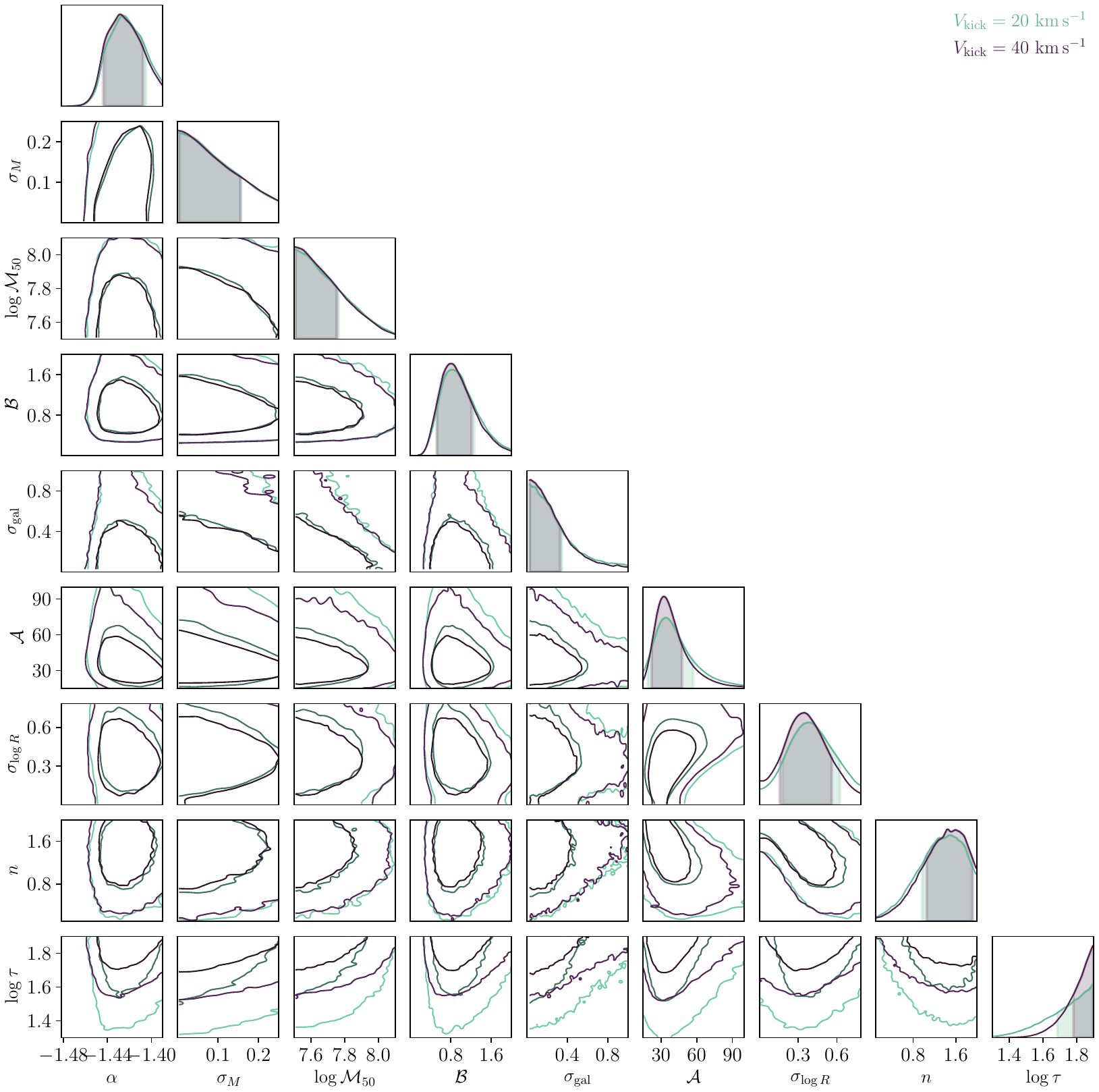}
\caption{Posterior distributions from our DDM fits to the DES and PS1 satellite populations for $V_{\rm kick}=20\kms$ (green) and $V_{\rm kick} = 40\kms$ (purple). The effects of dark matter decays are more pronounced due to the greater kick velocity in the $V_{\rm kick} = 40\kms$ case, raising the lower bound on the decay lifetime relative to the $V_{\rm kick} = 20\kms$ case. Constraints on the eight galaxy--halo connection parameters are consistent for both values of $V_{\mathrm{kick}}$. These parameters govern the faint-end slope ($\alpha$) and scatter ($\sigma_M$) of the satellite abundance-matching relation, the peak subhalo mass at which $50\%$ of subhalos host galaxies ($\log\mathcal{M}_{50}$, in units of $\Msun$), the efficiency of subhalo disruption due to the Galactic disk ($\mathcal{B}$), the scatter in the galaxy occupation fraction ($\sigma_{\mathrm{gal}}$), the amplitude ($\mathcal{A}$), scatter ($\sigma_{\log R}$), and power-law index ($n$) of the satellite--subhalo size relation, and the dark matter particle lifetime ($\log \tau$, in units of gigayears). Note that $\sigma_M$, $\sigma_{\rm{gal}}$, and $\sigma_{\log R}$ are reported in $\rm{dex}$ and $\mathcal{A}$ is reported in parsecs. Definitions for each parameter are given in Appendix \ref{sec:appendix_d}.}
\label{fig:corner}
\end{figure*}

Following \cite{PaperII,PaperIII}, we use the sample of kinematically confirmed and probable dwarfs from \cite{PaperI}. We remind the reader that, unlike in the WDM-like analyses from \cite{PaperIII}, we additionally assume that observed MW satellites occupy subhalos above a minimum present-day mass threshold of $\sim 2\times 10^7\Msun$, which was imposed when deriving our DDM SHMF suppression predictions. Although subhalos stripped below this resolution threshold are technically included in our analysis through our orphan satellite model following \cite{PaperII}, the abundance of these systems is assumed to follow an extrapolation of the DDM SHMF suppression derived in Section \ref{sec:ddm-sims}, and their contribution to our results is therefore negligible.

We use the Markov Chain Monte Carlo code \emcee \citep{1202.3665} to simultaneously fit for eight parameters governing the galaxy--halo connection and the efficiency of subhalo disruption due to the Galactic disk, and one parameter governing the impact of DDM.
In particular, we fit for the power-law slope of the satellite luminosity function ($\alpha$), the scatter in luminosity at fixed $V_{\rm peak}$ ($\sigma_M$), the mass at which 50\% of halos host galaxies ($\log(\mathcal{M}_{50}/\Msun)$), the strength of subhalo disruption due to baryons ($\mathcal{B}$), the scatter in the galaxy occupation fraction ($\sigma_{\mathrm{gal}}$), the amplitude of the galaxy--halo size relation ($\mathcal{A}$), the scatter in half-light radius at fixed halo size ($\sigma_{\log R}$), the power-law index of the galaxy--halo size relation ($n$), and the lifetime of the DM particle ($\log(\tau/\Gyr)$).
Note that we fit for $\log\tau$ (where $\tau$ is measured in gigayears), rather than $\tau$ itself, using a uniform prior on the logarithmic quantity in the range $\log \tau \in [1.3,1.9]$; the remaining prior distributions are identical to those adopted in \cite{PaperII}.\footnote{Although our DDM simulations sample down to $\tau = 10\Gyr$, we use a slightly higher lower-limit of the $\log\tau$ prior, below which the marginalized posterior is flat and nearly zero; this is a conservative choice.} We perform two separate fits at fixed $V_{\mathrm{kick}}=20\kms$ and $40 \kms$, each of which uses $36$ walkers, discards a burn-in period of $\sim 20$ autocorrelation lengths, and retains $\sim 10^5$ samples corresponding to $\sim 100$ autocorrelation lengths.

\subsection{Results}
\label{sec:results}

\begin{figure*}[t!]
\includegraphics[width=\textwidth]{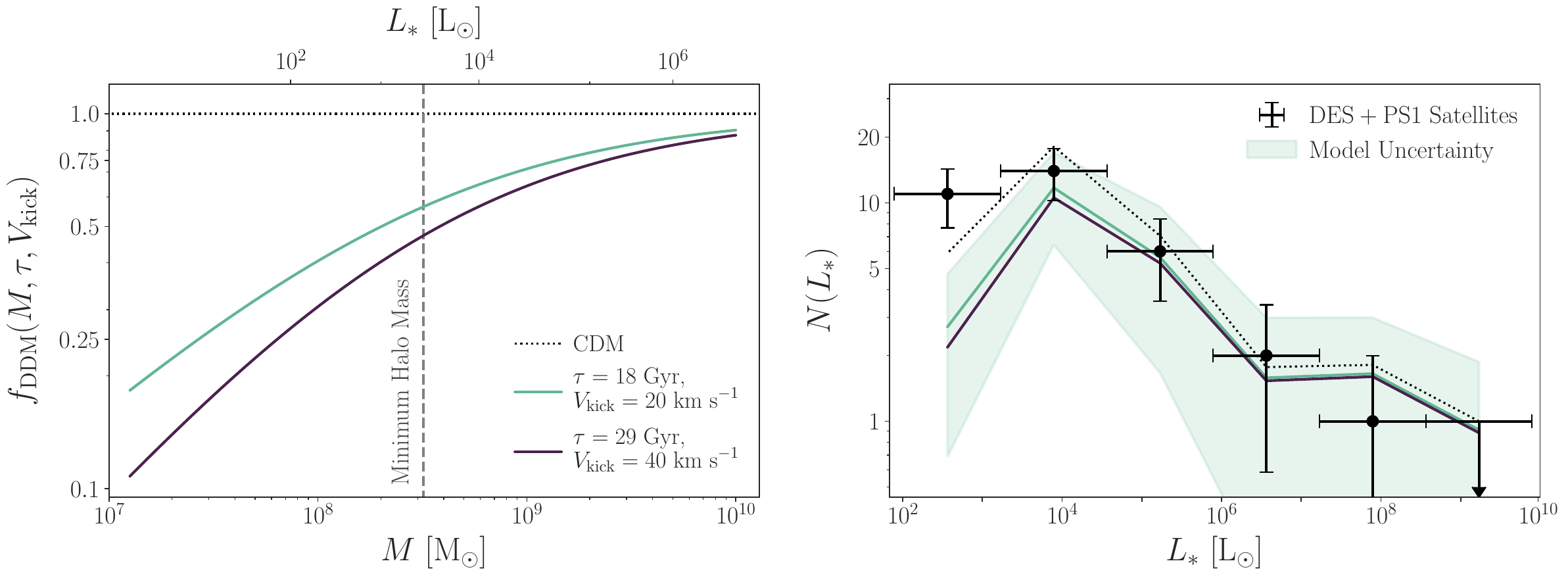}
    \caption{
        Predictions for the DDM models disfavored at the 95\% confidence level; these models represent the parameters for DDM sufficient to predict significant observational differences relative to observations.
        \emph{Left panel}: SHMF suppression relative to CDM for the disfavored DDM models. The vertical dashed line indicates the 95\% confidence upper limit on the lowest-mass halo inferred to host MW satellite galaxies \citep{PaperII}.
        \emph{Right panel}: predicted MW satellite galaxy luminosity functions for the disfavored DDM models compared to DES and PS1 observations, evaluated at the best-fit MW satellite model parameters from \citet{PaperII}. The shaded band illustrates the uncertainty of the DDM prediction due to the stochasticity of our galaxy--halo connection model and the limited number of simulations used in our analysis; the size of the uncertainty is very similar to that in CDM. Note that this panel is a simple one-dimensional representation of our MW satellite and DM model fit to the luminosity, size, and spatial distribution of satellites in the DES and PS1 survey footprints. The comparison of our CDM model to data is described in \citet{PaperII}.}
\label{fig:prediction}
\end{figure*}

\begin{table*}[ht!]
    \caption{Prior distributions and 95\% Credible Intervals Derived from the Marginalized Posterior for Each Parameter from Our DDM Fits to the MW Satellite Population for $V_{\rm kick}=20\kms$ and $V_{\rm kick}=40\kms$ (\figref{corner}).}
    \begin{tabular*}{\textwidth}{r | l @{\extracolsep{\fill}} l l}
        \hline\hline
        \multirow{2}*{Parameter} & \multirow{2}*{Prior Distribution} & \multicolumn{2}{c}{95\% Credible Interval} \\
        \hfill & \hfill & $V_{\rm kick} = 20\kms$ & $V_{\rm kick} = 40\kms$ \\
        \hline
        $\alpha$ & $\arctan\alpha \sim \operatorname{unif}(-1.1,-0.9)$ & $[-1.452, -1.378]$ & $[-1.454, -1.381]$ \\
        $\sigma_M$ & $\sigma_M \sim \operatorname{unif}(0,2)\dex$ & $[0.00^*, 0.36]\dex$ & $[0.00^*, 0.37]\dex$ \\
        $\log(\mathcal{M}_{50}/\Msun)$ & $\log(\mathcal{M}_{50}/\Msun) \sim \operatorname{unif}(7.5,11.0)$ & $[7.50^*, 8.01]$ & $[7.50^*, 7.99]$ \\
        $\mathcal{B}$ & $\ln\mathcal{B} \sim \mathcal{N}(\mu=1.0,\sigma=0.5)$ & $[0.33, 1.84]$ & $[0.33, 1.75]$ \\
        $\sigma_{\mathrm{gal}}$ & $\sigma_{\mathrm{gal}} \sim \operatorname{unif}(0,1)\dex$ & $[0.03^*, 0.79]\dex$ & $[0.03^*, 0.76]\dex$ \\
        $\mathcal{A}$ & $\mathcal{A} \sim \operatorname{unif}(0.0,0.5)\kpc$ & $[15, 93]\pc$ & $[16, 76]\pc$ \\
        $\sigma_{\log R}$ & $\sigma_{\log R} \sim \operatorname{unif}(0,2)\dex$ & $[0.01^*, 0.86]\dex$ & $[0.00^*, 0.75]\dex$ \\
        $n$ & $n \sim \mathcal{N}(\mu=1.0,\sigma=0.5)$ & $[0.50, 2.00^*]$ & $[0.56, 2.00^*]$ \\
        $\log(\tau/\Gyr)$ & $\log(\tau/\Gyr) \sim \operatorname{unif}(1.3,1.9)$ & $[1.46, 1.90^*]$ &$[1.63, 1.90^*]$ \\
        \hline
    \end{tabular*}
    \tablecomments{The first eight parameters describe the galaxy--halo connection model used to associate satellite galaxies with subhalos in our MW-like zoom-in simulations \citep{PaperII}, and the final parameter corresponds to the DDM particle lifetime. Definitions for each parameter are given in Appendix \ref{sec:appendix_d}. Asterisks denote prior-constrained limits. See Table 2 of \citet{PaperII} for motivations for the prior distributions of the eight parameters describing the galaxy--halo connection model; note that the prior on $n$ is bounded between 0.0 and 2.0 for convergence.}
    \label{tab:parameters}
\end{table*}

The posterior distributions from our $V_{\rm kick}=20\kms$ and $V_{\rm kick}=40\kms$ fits are summarized in Table \ref{tab:parameters} and shown in \figref{corner}. The marginalized posterior distributions for the eight galaxy--halo connection parameters are nearly identical in both cases, and are consistent with the CDM fit in \cite{PaperII}. For both values of $V_{\mathrm{kick}}$, we obtain a lower limit on $\log\tau$, which is expected because the CDM model (i.e., the large $\tau$ limit) is consistent with the data. At 95\% confidence, we obtain $\log(\tau/\Gyr) > 1.46$ ($\tau > 29\Gyr$) for $V_{\mathrm{kick}}=20\kms$ and $\log(\tau/\Gyr) > 1.63$ ($\tau > 43\Gyr$) for $V_{\mathrm{kick}}=40\kms$ (Table \ref{tab:parameters}).

Following \cite{PaperIII}, we scale these constraints to conservatively account for uncertainty in the MW host halo mass, which is known to impact limits on non-CDM models derived from the MW satellite population (e.g., \citealt{Newton201108865}). In particular, for each $V_{\mathrm{kick}}$, we compute the value of $\tau$ that decreases $f_{\mathrm{DDM}}$ by $27\%$ relative to its value at the original $\tau$ constraint when evaluated at a peak subhalo mass of $3.2\times 10^8 \Msun$. This value corresponds to the minimum halo mass probed by the DES and PS1 data in CDM \citep{PaperII}, and therefore represents an upper limit on the minimum halo mass in our DDM inference. The $27\%$ uncertainty corresponds to the ratio of the maximum allowed MW halo mass from \cite{Callingham180810456}, appropriately converted to our virial mass definition, relative to the average host halo mass from our MW-like simulations used to perform the inference. We assume that this uncertainty affects the allowed amount of SHMF suppression linearly when deriving our conservative constraints on $\tau$.
For lower values of $\tau$, the number of MW satellites with $L \lesssim 10^4\ L_{\mathrm{\odot}}$ predicted to be observed in the DES and PS1 footprints is significantly lower than the data, similar to the alternative DM models shown in Figure 1 of \cite{PaperIII}.

This procedure yields our fiducial and conservative $95\%$ confidence constraints of $\tau> 18\Gyr$ for $V_{\mathrm{kick}}=20\kms$ and $\tau> 29\Gyr$ for $V_{\mathrm{kick}}=40\kms$.
The SHMF suppression relative to CDM and the predicted luminosity function of DES and PS1 satellites for each of these models are shown in the left and right panels of \figref{prediction}, respectively. The right panel of \figref{prediction} also compares these predictions to the observed luminosity function, demonstrating that the DDM models that our analysis rules out yield significantly fewer ultra-faint satellites than observed by DES and PS1, after accounting for observational selection effects and conservatively marginalizing over modeling uncertainties.

\begin{figure}
\includegraphics[width=\columnwidth]{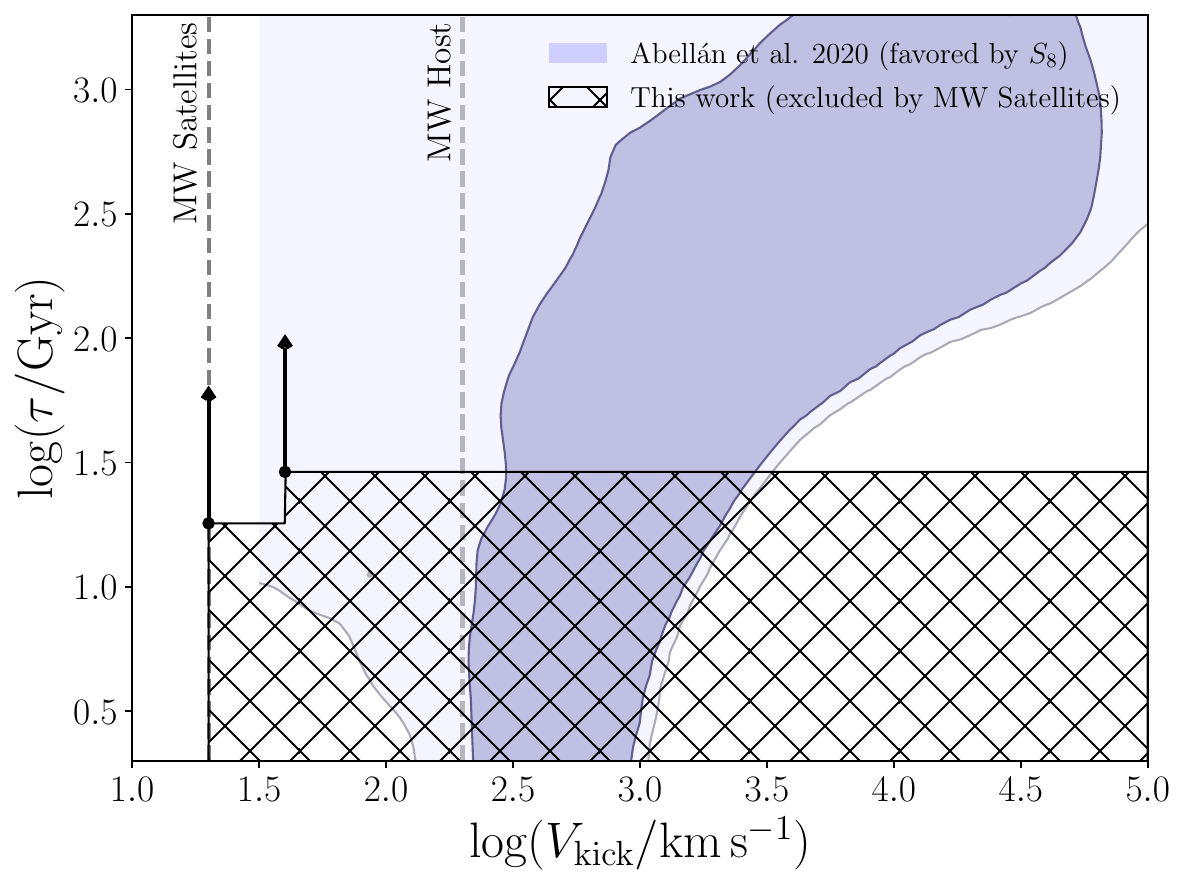}
    \caption{Lower limits from our fits for $V_{\rm kick} = 20\kms, 40\kms$ are shown as points with black arrows indicating allowed parameter space. The corresponding excluded region is identified with black crosshatching. The region of DDM parameter space that potentially alleviates the $S_8$ tension \citep{2008.09615} is shown by the lavender contour, where dark (light) shading shows $68\%$ ($95\%$) confidence intervals. Note that the posterior distribution from \cite{2008.09615} is limited by their prior range and does not extend to values of $V_{\rm kick}$ as low as those studied in this paper. The typical magnitudes of peak velocities for MW satellite halos and host halo are indicated by vertical dashed gray lines.}
\label{fig:constraint}
\end{figure}

\figref{constraint} shows these lower limits on $\log(\tau/\Gyr)$ for $V_{\rm kick} = 20\kms$ and $40\kms$ alongside the preferred region of DDM parameter space that potentially alleviates the $S_8$ tension \citep{2008.09615}. As discussed in detail in Section \ref{sec:comparison}, these constraints very conservatively exclude roughly half of the DDM parameter space favored to resolve $S_8$ tension \citep{2008.09615} and the $H_0$ tension \citep{1903.06220}. Our results are conservative in this context because our fits are performed at extremely low $V_{\mathrm{kick}}$ relative to the typical values used to alleviate these cosmological tensions. Moreover, our fit only directly incorporates the effects of DDM physics on subhalo and satellite abundances, and therefore does not leverage the reduced central densities of DDM subhalos, which may yield additional constraining power.

\section{Discussion}
\label{sec:context}

In this section, we place our DDM constraints in the context of previous studies (Section \ref{sec:comparison}) and compare our constraints to those derived for other dark matter models using a similar analysis of the MW satellite population (Section \ref{sec:models}).

\subsection{Comparison to Previous Studies}
\label{sec:comparison}

We now place our results in the context of previous DDM studies, including (but not limited to) studies of MW satellite galaxies. We reiterate that our analysis makes conservative assumptions regarding both the microphysics of DM decays and their impact on structure formation. In particular, our constraints only directly apply to two-body decays that yield a cold, stable DM daughter particle, and to DDM models with sufficiently long lifetimes such that small-scale structure is only significantly affected relative to CDM at late times. We therefore caution that it is not straightforward to compare our results with limits from DDM models with different decay mechanisms or to limits on short-lifetime decays that affect the pre-recombination universe. As a result, the following discussion focuses on limits derived for the same family of DDM models considered in our analysis.
Beyond these cases, mapping the suppression of the subhalo mass function for our fiducial two-body decay model to that in a single-body decay would allow for a more direct comparison with many large-scale structure analyses \citep[e.g.,][]{2011.04606,Hubert210407675}; this is left to a future work.

\subsubsection{Limits from Milky Way Satellites}

Several authors have studied the impact of decays on MW satellite galaxies for the same class of DDM models we consider. In particular, \citet{1009.1912} compared the population of classical and Sloan Digital Sky Survey-discovered MW satellite galaxies to DDM predictions. These authors found that $\tau \lesssim 30\Gyr$ is ruled out for~$20\kms \lesssim V_{\rm kick} \lesssim 200 \kms$, where specific results depend on assumptions regarding the star-formation histories of satellites and the evolution of subhalos and satellites in their semi-analytic model in detail.\footnote{\citet{1011.4970} synthesized the results of \citet{1003.0419} and \citet{1009.1912}, reporting that $\tau \lesssim 40\Gyr$ is ruled out for $V_{\rm kick}>20\kms$.} These limits are consistent with and slightly weaker than our fiducial constraints.

The most stringent limits from \citet{1009.1912} are driven by the inferred mass enclosed within $300\pc$ of each MW satellite rather than the overall abundance of these systems. Thus, these limits are mainly set by the mass loss and reduction in central densities of DDM subhalos, rather than their enhanced disruption relative to CDM. From an observational standpoint, the enclosed mass depends on the measured stellar velocity dispersion for each satellite and a mapping between this quantity and a halo mass proxy (e.g., $V_{\rm max}$), both of which are accompanied by significant systematic uncertainties.

On the other hand, our DDM constraints are driven by the abundance of confirmed and candidate dwarf galaxies in the DES and PS1 datasets as a function of absolute magnitude, half-light radius, and heliocentric distance, which we leverage in a statistical forward model. Integrating comparisons of predicted and inferred central dynamical masses into our model is an important area for future work that will likely yield even more stringent constraints on DDM.

Several other authors have studied whether the family of DDM models that we analyze can reconcile the apparent tension between the predicted and inferred density profiles of dwarf galaxy halos. In particular, \citet{Wang:2014} found that lifetimes $\tau\sim\mathcal{O}(10\Gyr)$ and kick velocities $V_{\rm kick}\sim 20\kms$ are consistent with classical MW satellite galaxies. Meanwhile, \citet{2011.13121} studied the density profile of isolated dwarf galaxies, finding that $\tau\lesssim 7.0 \Gyr$ is needed to explain these measurements for $V_{\rm kick}= 20\kms$; these results are consistent with those in \cite{0305496} and \cite{0806.0602}. 

Comparing our constraints to these previous results implies that DDM models that significantly alter dwarf galaxy central densities are not viable because they simultaneously reduce MW satellite galaxy abundances to unacceptable levels. Similar conclusions have been drawn for warm and fuzzy DM, which drastically suppress the abundance of ultra-faint dwarf galaxies before they yield an observable impact on the density profiles of brighter dwarfs \citep{Maccio12021282,Safazadeh190611848,PaperIII}.

\subsubsection{Limits from Galaxy Clusters}

\citet{1003.0419} used the halo mass--concentration relation and mass function derived from galaxy clusters to estimate limits on the DDM kick velocity and lifetime, finding that galaxy cluster observations rule out decay times less than a few times the age of the universe for kick velocities greater than $\sim 100\kms$. These results are qualitatively consistent with our constraints in \figref{constraint}, in the sense that larger values of $V_{\rm kick}$ result in more stringent constraints on $\tau$, and they disfavor regions of parameter space that lie along this degeneracy at larger $V_{\rm kick}$ than sampled by our analysis. As noted by \citet{1003.0419}, these results are approximate and warrant a detailed statistical analysis based on cosmological DDM simulations.

\subsubsection{Limits from the Lyman-$\alpha$ Forest}

Aside from the impact of DDM microphysics on low-mass subhalos at late times, DM decays can leave a substantial imprint on small-scale structure throughout cosmic history. For example, \citet{1309.7354} used observations of the Lyman-$\alpha$ forest to exclude $\tau \lesssim 10\Gyr$ and $V_{\rm kick} \gtrsim 30$--$70\kms$ for the same class of DDM models we consider. These constraints are again consistent with, and weaker than, our fiducial results, which is reasonable given that recent MW satellite constraints on WDM are more stringent than WDM constraints derived using the Lyman-$\alpha$ forest data considered in \citealt{1309.7354} (e.g., see \citealt{PaperIII}). As discussed in \citet{1309.7354}, more precise limits likely entail a joint inference of intergalactic medium (IGM) properties and small-scale clustering based on hydrodynamic DDM simulations because IGM properties are partially degenerate with DM properties (e.g., \citealt{GarzilliMNRAS}).

We note that low-redshift tracers of small-scale structure, including observations of ultra-faint dwarf galaxies, stellar streams, and strong gravitational lenses, are well suited to test DDM physics because the impact of DM decays on small-scale structure becomes more severe at late times.
Thus, a detailed study of the synergies between small-scale structure probes that sample distinct scales and redshifts, including ultra-faint dwarf galaxies, stellar streams, strong gravitational lensing, and the Lyman-$\alpha$ forest, is particularly interesting in the context of DDM and promises to yield precise joint measurements (see \citealt{Enzi201013802,Nadler210107810} for examples of joint WDM constraints).

\subsubsection{Limits from Other Cosmological Probes}

DDM has recently gained interest as a potential solution to the $H_0$ and $\sigma_8$ tensions because it can potentially reduce the late-time expansion rate and matter power spectrum without strongly affecting early universe observables including the CMB. In particular, \citet{1903.06220} found that DDM can resolve the $H_0$ tension for $\tau = 35\Gyr$ and $\epsilon = 0.16$ ($V_{\rm kick} \approx 10^4 \kms$) by combining late-time measurements of the expansion rate including distance-ladder, baryonic acoustic oscillation (BAO), quasar, and Lyman-$\alpha$ auto- and cross-correlation data. However, subsequent analyses have shown that this model is excluded by additional datasets. Specifically, \citet{2004.07709} used Type-Ia Supernovae, BAO, and time delay measurements of gravitationally lensed quasars with priors set by the CMB to derive a limit of $\tau > 9\Gyr$ for two-body decays at the maximum allowed value of $\epsilon=0.5$ ($V_{\rm kick} \approx 10^5\kms$) in their analysis. Similarly, \citet{2006.03678} used \emph{Planck} CMB power spectra measurements to derive $\tau>1000\Gyr$ for $\epsilon=0.5$. Neither of these analyses found that viable DDM models (at such high mass splitting ratios) can significantly affect the present-day expansion rate.

Meanwhile, \citet{2008.09615} found that DDM can resolve the $S_8$ tension for $\tau = 56\Gyr$ and $\epsilon \approx 0.007$ ($V_{\rm kick} \approx 10^3\kms$) by combining \emph{Planck} CMB lensing and power spectra, BAO data, and Type-Ia Supernovae with KIDS1000+BOSS+2dfLenS measurements of $S_8$. \citet{Abellan210212498} found that this conclusion is robust to the inclusion of additional cosmological and experimental constraints, and that it can potentially explain the anomalous Xenon1T electron recoil excess \citep{Aprile200609721,Kannike200610735}.

Our analysis excludes $\tau < 18\Gyr$ for $V_{\rm kick} = 20\kms$ ($\epsilon \approx 10^{-4}$); this value of $V_{\rm kick}$ is significantly lower than those preferred by the $H_0$ and $S_8$ analyses described above. Because the impact of DDM on small-scale structure becomes more severe for larger values of $V_{\rm kick}$, we can extremely conservatively interpret this fiducial constraint on $\tau$ as a limit on the DDM models considered by, e.g., \citet{1903.06220} and \citet{2008.09615}.
As shown by the crosshatched regions in \figref{constraint}, this disfavors the preferred DDM parameter space reported by \citet{2008.09615} for low values of $\tau$ and $V_{\mathrm{kick}}$.\footnote{Note that the posterior from \citet{2008.09615} is prior-limited and not strongly constrained as a function of $\tau$.} In practice, we expect our constraints on $\tau$ to become much more stringent for these models. For example, because halos with characteristic virial velocities of $\mathcal{O}(V_{\rm kick})$ are affected by DM decays, we expect that the $V_{\rm kick}\approx 10^3\kms$ model considered by \citet{2008.09615} would significantly affect the structure and abundance of halos that host MW-mass and even larger galaxies. In turn, structure on dwarf galaxy scales is likely completely different than in CDM (and therefore incompatible with MW satellite data) for all values of $\tau$ preferred by these analyses, though dedicated future work is necessary to quantify this claim.

\subsection{Comparison to Constraints on Other Dark Matter Models from Milky Way Satellites}
\label{sec:models}

We now place our DDM results in the context of constraints on other DM models derived from the MW satellite population and discuss implications for small-scale structure observables.
We limit our quantitative comparison to the WDM constraints from \cite{PaperIII} because this study used an identical modeling framework with the exception of the assumed SHMF suppression; however, we note that many other studies have used the MW satellite population to derive WDM constraints (e.g., \citealt{Maccio09102460,Polisensky10041459,Anderhalden12122967,Kennedy13107739,Kim171106267,Nadler190410000,Newton201108865,Dekker211113137}).\footnote{\citet{PaperIII} also derived constraints on DM--baryon interactions and fuzzy DM based on the suppression of the linear matter power spectrum and low-mass halo abundances in these models. We consider these ``WDM-like'' models for this discussion, although they may have distinct effects on the MW satellite population in detail.} 

\cite{PaperIII} found that thermal relic WDM masses below $6.5\ \mathrm{keV}$ are excluded by the same MW satellite census that we consider at $95\%$ confidence after marginalizing over an identical set of galaxy--halo connection and MW halo mass parameters. This WDM model suppresses subhalo abundances by $\sim 25\%$ relative to CDM at the minimum observed halo mass scale of $3.2\times 10^8\Msun$; the SHMF declines rapidly at smaller masses, reaching $\sim 75\%$ suppression relative to CDM at its half-mode mass of $3.8\times 10^7\Msun$. For comparison, the DDM models we rule out at $95\%$ confidence suppress the subhalo mass function by $\sim 50\%$ at the minimum observed halo mass scale, and their SHMFs decline less rapidly than for the ruled-out WDM model at lower peak subhalo masses. We reiterate that our fit to the DDM SHMF suppression is only valid for subhalos above a present-day mass threshold of $\sim 2\times 10^7\Msun$, and that robust estimates for subhalo abundances at lower masses require higher-resolution simulations.

Although regions of parameter space for both WDM-like and DDM models are excluded by the abundance of known MW satellites, these scenarios make distinct predictions for other dwarf galaxy and small-scale structure observables. In particular, decays can significantly deplete both low-mass isolated halos and subhalos of dark matter at late times, lowering predicted mass-to-light ratios for both satellite and field dwarf galaxies relative to CDM. Similarly, mass loss and momentum transfer due to decays reduce halos' central densities, which future observations of dynamical tracers in MW satellites will better inform \citep{Simon190304743}. On the other hand, WDM halos with masses well above the half-mode scale do not significantly differ in present-day mass relative to CDM, though delayed formation lowers their concentration \citep{Bose150701998,Stucker210909760}. Combining our constraints with small-scale structure observables that are sensitive to the abundance and internal structure of halos and subhalos at late times, including strong gravitational lensing (e.g., \citealt{Minor161205250,Hsueh190504182,Gilman190806983,Gilman190902573}), will therefore help differentiate these classes of models.

In addition to the WDM-like models discussed above, it is also interesting to contrast the effects of DDM with those of self-interacting DM (SIDM), which can also suppress the abundance of low-mass subhalos while altering their density profiles \citep{Vogelsberger12015892,Zavala12116426,Tulin170502358,Robles190301469,Nadler200108754,Nadler210912120}. Unlike DDM, which (to first order) equally depletes both isolated halos and subhalos of dark matter relative to CDM, subhalos' mass loss and disruption in SIDM are closely tied to their orbital histories (e.g., \citealt{Dooley160308919,Jiang210803243}). Thus, comparing the abundance and internal structure of isolated halos and subhalos---for example, through joint analyses of field and satellite dwarf galaxy populations or of line-of-sight and subhalo perturbations in strong lensing data---will help disentangle these forms of dynamical DM microphysics. These differences can also potentially be tested by comparing the abundance and properties of LMC-associated MW satellites with the remainder of the MW satellite population \citep{Nadler210912120}.


\section{Conclusions}
\label{sec:conclusion}

We have used a state-of-the-art census of the MW satellite galaxy population to set robust and stringent constraints on the DM particle lifetime that are among the most robust and stringent to date while making conservative assumptions about the decay mechanism (i.e., late-time decays that include a stable CDM-like daughter product). In particular, we combined cosmological zoom-in simulations of DDM with a forward model of the MW satellite population to jointly infer the connection between these galaxies and their DM subhalos and the potential impact of DM decays on the abundance of these systems.

For DM that undergoes late-time two-body decays to a massless dark radiation species and a cold, stable daughter DM particle, we find that:
\begin{enumerate}
    \item For DM particle lifetimes of $\mathcal{O}(t_H)$ and recoil kick velocities of $\mathcal{O}(10\kms)$, the smallest subhalos that host ultra-faint dwarf galaxies ($M_{\rm peak} \sim 10^8 \Msun$) can lose a significant fraction of their peak mass due to decays alone;
    \item DM decays can suppress the abundance of surviving subhalos above a minimum mass threshold of $\approx 4\times 10^7\Msun$ at the $\sim 50\%$ level relative to CDM. This suppression is approximately independent of Galactocentric radius, and \eqnref{f_ddm} provides a fitting function for this effect derived from cosmological zoom-in simulations;
    \item The population of MW satellite galaxies observed by DES and PS1 excludes DDM models with decay lifetime $\tau < 18\Gyr$ ($29\Gyr$) for $V_{\mathrm{kick}}=20\kms$ ($40\kms$) at $95\%$ confidence;
    \item These constraints can be conservatively extrapolated to higher $V_{\rm kick}$ values to exclude approximately half of the DDM parameter space preferred to alleviate the $H_0$ and $S_8$ tensions. These constraints are expected to become more stringent for the $V_{\rm kick}$ values considered in those analyses and with the inclusion of MW satellite stellar velocity dispersion measurements;
    \item Combining our DDM constraints based on MW satellites with complementary small-scale structure probes at low and high redshifts---including field dwarf galaxy luminosity functions, strong gravitational lensing, and the Lyman-$\alpha$ forest---will help differentiate the effects of decays from other DM microphysics.
\end{enumerate}

Our analysis only directly leverages the reduction of DDM subhalo abundances and its impact on MW satellite abundances, rather than the (potentially observable) effects on the internal dynamics of satellite galaxies. Nonetheless, our results are consistent with and more stringent than previous limits driven by MW satellite stellar velocity dispersion measurements. Future work that combines our approach with the inferred dynamical masses and density profiles of MW satellites promises to further improve DDM constraints, as do future detections of dwarf galaxies within and beyond the MW. These observational advances will, respectively, be enabled by forthcoming spectroscopic facilities and giant segmented mirror telescopes \citep{Simon190304743} and observational facilities including the Vera C.\ Rubin Observatory \citep{Drlica-Wagner:2019,Mutlu-Pakdil210501658} and the Nancy Grace Roman Space Telescope.

Although other cosmological observables disfavor DDM as a solution to the $H_0$ and $S_8$ tensions (e.g., \citealt{2004.07709,2006.03678}),\footnote{We note that \citealt{2006.03678} approximated the perturbations of the massive daughter particles, which were found to be essential to account for the effects on the $S_8$ parameter by \citealt{2008.09615}.} we emphasize that our results inform DDM physics as a solution to these tensions in a way that is complementary to expansion history and large-scale structure probes.
In particular, large-scale structure probes are primarily sensitive to the DM lifetime, regardless of the microphysical decay mechanism, while the abundance of low-mass DM subhalos traced by MW satellite galaxies is sensitive to both the lifetime of the DM particle and the decay mechanism as encapsulated by $V_{\rm kick}$ in our model.
Thus, to more fully inform DDM physics, it is crucial to combine observables that cover a wide range of cosmological epochs and scales, including probes of small-scale structure.


\section{Acknowledgments}

The authors thank Guillermo Abell\'an for sharing their results from \citet{2008.09615} and Andrew Benson for comments on the manuscript.

This research received support from the National Science Foundation (NSF) under grant No.\ NSF DGE-1656518 through the NSF Graduate Research Fellowship received by S.M.\ and E.O.N.\ and from the U.S. Department of Energy under contract number DE-AC02-76SF00515 to SLAC National Accelerator Laboratory. Y.-Y.M.\ is supported by NASA through the NASA Hubble Fellowship grant No.\ HST-HF2-51441.001 awarded by the Space Telescope Science Institute, which is operated by the Association of Universities for Research in Astronomy, Inc., under NASA contract NAS5-26555.

This research made use of computational resources at SLAC National Accelerator Laboratory, a U.S.\ Department of Energy Office; the authors are thankful for the support of the SLAC computational team. 
This research made use of the Sherlock
cluster at the Stanford Research Computing Center (SRCC); the authors are thankful for the support of the SRCC team. 
This research made use of arXiv.org (\url{https://arXiv.org}) and NASA's Astrophysics Data System for bibliographic information.
This research made use of adstex (\url{https://github.com/yymao/adstex}).
This research made use of the cubehelix color scheme \citep{Green:2011}.

Funding for the DES Projects has been provided by the U.S. Department of Energy, the U.S. National Science Foundation, the Ministry of Science and Education of Spain, 
the Science and Technology Facilities Council of the United Kingdom, the Higher Education Funding Council for England, the National Center for Supercomputing 
Applications at the University of Illinois at Urbana-Champaign, the Kavli Institute of Cosmological Physics at the University of Chicago, 
the Center for Cosmology and Astro-Particle Physics at the Ohio State University,
the Mitchell Institute for Fundamental Physics and Astronomy at Texas A\&M University, Financiadora de Estudos e Projetos, 
Funda{\c c}{\~a}o Carlos Chagas Filho de Amparo {\`a} Pesquisa do Estado do Rio de Janeiro, Conselho Nacional de Desenvolvimento Cient{\'i}fico e Tecnol{\'o}gico and 
the Minist{\'e}rio da Ci{\^e}ncia, Tecnologia e Inova{\c c}{\~a}o, the Deutsche Forschungsgemeinschaft and the Collaborating Institutions in the Dark Energy Survey. 

The Collaborating Institutions are Argonne National Laboratory, the University of California at Santa Cruz, the University of Cambridge, Centro de Investigaciones Energ{\'e}ticas, 
Medioambientales y Tecnol{\'o}gicas-Madrid, the University of Chicago, University College London, the DES-Brazil Consortium, the University of Edinburgh, 
the Eidgen{\"o}ssische Technische Hochschule (ETH) Z{\"u}rich, 
Fermi National Accelerator Laboratory, the University of Illinois at Urbana-Champaign, the Institut de Ci{\`e}ncies de l'Espai (IEEC/CSIC), 
the Institut de F{\'i}sica d'Altes Energies, Lawrence Berkeley National Laboratory, the Ludwig-Maximilians Universit{\"a}t M{\"u}nchen and the associated Excellence Cluster Universe, 
the University of Michigan, NSF's NOIRLab, the University of Nottingham, The Ohio State University, the University of Pennsylvania, the University of Portsmouth, 
SLAC National Accelerator Laboratory, Stanford University, the University of Sussex, Texas A\&M University, and the OzDES Membership Consortium.

Based in part on observations at Cerro Tololo Inter-American Observatory at NSF's NOIRLab (NOIRLab Prop. ID 2012B-0001; PI: J. Frieman), which is managed by the Association of Universities for Research in Astronomy (AURA) under a cooperative agreement with the National Science Foundation.

The DES data management system is supported by the National Science Foundation under Grant Numbers AST-1138766 and AST-1536171.
The DES participants from Spanish institutions are partially supported by MICINN under grants ESP2017-89838, PGC2018-094773, PGC2018-102021, SEV-2016-0588, SEV-2016-0597, and MDM-2015-0509, some of which include ERDF funds from the European Union. IFAE is partially funded by the CERCA program of the Generalitat de Catalunya.
Research leading to these results has received funding from the European Research
Council under the European Union's Seventh Framework Program (FP7/2007-2013) including ERC grant agreements 240672, 291329, and 306478.
We  acknowledge support from the Brazilian Instituto Nacional de Ci\^encia
e Tecnologia (INCT) do e-Universo (CNPq grant 465376/2014-2).

This manuscript has been authored by Fermi Research Alliance, LLC under Contract No. DE-AC02-07CH11359 with the U.S. Department of Energy, Office of Science, Office of High Energy Physics.

\software{
    \code{NumPy} \citep{harris2020array},
    \code{SciPy} \citep{2020SciPy-NMeth},
    \code{Matplotlib} \citep{Hunter:2007},
    \code{seaborn} \citep{Waskom2021},
    \code{emcee} \citep{1202.3665},
    \code{ChainConsumer},\footnote{\url{https://github.com/Samreay/ChainConsumer}}
    \code{incredible},\footnote{\url{https://github.com/abmantz/incredible}}
    \code{GADGET} \citep{Springel:2005},
    \code{ROCKSTAR} \citep{1110.4372},
    \code{CONSISTENT-TREES} \citep{1110.4370}.
}


\bibliography{main}


\appendix

\section{Milky Way-like Simulations}
\label{sec:appendix_a}

To perform the inference in Section \ref{sec:constraints}, we use cosmological zoom-in simulations of two MW-like halos originally presented in \cite{Mao:2015} and studied in \citet{PaperII,PaperIII} to analyze the MW satellite population. 
Furthermore, to validate our DDM subhalo population predictions derived from the expanded \cite{Wang:2014} simulation suite, we perform DDM resimulations of these MW-like systems.
These host halos are selected based on mass and concentration estimates for the MW and due to the presence of realistic LMC analogs and early \textit{Gaia}-Sausage-Enceladus-like merger events with the properties described in \cite{PaperII}.
We simulate both of these systems in DDM models with $(\tau/\Gyr, V_{\rm kick}/\kms) \in \{(20,20),(40,30)\}$, roughly corresponding to the 95\% confidence level constraints determined in \secref{constraints}.

To perform the DDM resimulations, we use the same modified version of the \code{GADGET-2} and \code{GADGET-3} N-body codes from \citet{1003.0419}. The original CDM MW-like simulations and the DDM resimulations are run with $\Omega_M=0.286$, $\Omega_\Lambda=0.714$, $n_s=1$, $h=0.7$, and $\sigma_8=0.82$ \citep{WMAP9}.\footnote{Note that the $n_s$ value is in fact $1$ instead of $0.96$ as stated in previous studies \citep[e.g.,][]{Mao:2015}.} The highest-resolution region is simulated with a Plummer-equivalent force softening of $170\pc\ h^{-1}$ and a particle mass of $3.0\times 10^5\Msun\ h^{-1}$. We analyze these simulations using the \code{ROCKSTAR} \citep{1110.4372} halo finder and \code{CONSISTENT-TREES} \citep{1110.4370} merger tree code.

We note that the differences in low-mass subhalo abundances introduced by changes to the numerical and cosmological parameters with respect to the \citet{Wang:2014} simulations are minor compared to the differences between the DDM and CDM simulations within each suite (see, e.g., \citealt{Dooley14036828} for a study of the impact of cosmological parameters on subhalo statistics). The largest difference in cosmological parameters is the value of $n_s$; however, once the orbital phase of the LMC is fixed, the SHMF in the $n_s=1$ simulations we use is enhanced by $\sim 10\%$ at all subhalo masses relevant for our study compared to simulations with $n_s=0.96$. This allows for more severe SHMF suppression due to DDM when comparing to the data, meaning that the constraints we derive are conservative.

\section{Subhalo Mass Function Suppression Validation}
\label{sec:appendix_b}

\begin{figure*}
\includegraphics[width=\textwidth]{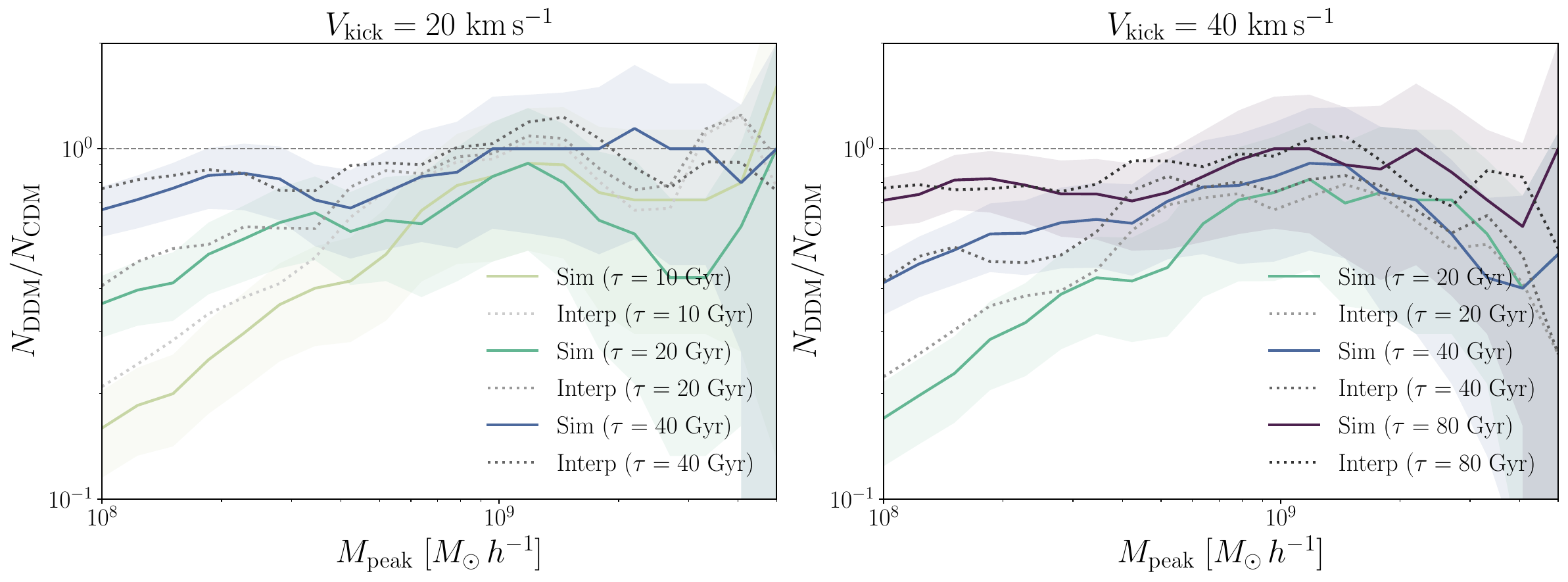}
\caption{SHMF suppression at $z=0$ for the MW-mass host halo in our expanded suite of zoom-in simulations based on \cite{Wang:2014} measured directly from the simulations (solid lines) and using the interpolating procedure described in Appendix \ref{sec:appendix_b} (dotted lines). Results are shown for DDM models with $\tau = 10$, $20$, and $40\Gyr$ (from yellow to purple), with $V_{\mathrm{kick}}=20\kms$ (left panel) and $40\kms$ (right panel) as a function of peak subhalo virial mass $M_{\mathrm{peak}}$. All SHMFs are restricted to subhalos above a conservative resolution threshold of $V_{\rm peak} > 10 \kms$ and $V_{\rm max} > 9 \kms$. Shaded bands indicate $68\%$ confidence interval Poisson uncertainties on the simulation measurements. The SHMF suppression predicted by our interpolating procedure is consistent with that measured directly from the simulations at the $\sim 1\sigma$ level.
}
\label{fig:sim-interp}
\end{figure*}

We use our MW-like resimulations to test that the impact of DDM on low-mass subhalos derived from the \citet{Wang:2014} simulations---and particularly the suppression of the DDM SHMF---is applicable to hosts of similar masses that specifically resemble the MW system. To do so, we compare predictions from ((i) \eqnref{f_ddm} and (ii) an interpolation of the SHMF suppression from the \cite{Wang:2014} simulations to the SHMF suppression determined directly from our MW-like resimulations. For the second comparison, we use piecewise linear interpolation to smoothly connect the SHMF suppression, $N(>M_{\rm peak})_{\rm DDM} / N(>M_{\rm peak})_{\rm CDM}$, between the points in DDM parameter space simulated by \citet{Wang:2014}: $(\tau/\Gyr, V_{\rm kick}/\kms) \in \{(10, 20),\allowbreak (20, 20),\allowbreak (20, 40),\allowbreak (40, 20),\allowbreak (40, 40),\allowbreak (80, 40)\}$.
This provides an alternative estimate of the SHMF suppression in regions of DDM parameter space that were not directly simulated.
The interpolating function is compared to the \citet{Wang:2014} simulation results in \figref{sim-interp}, which demonstrates agreement at the $\sim 1\sigma$ level.

\begin{figure*}
\includegraphics[width=\textwidth]{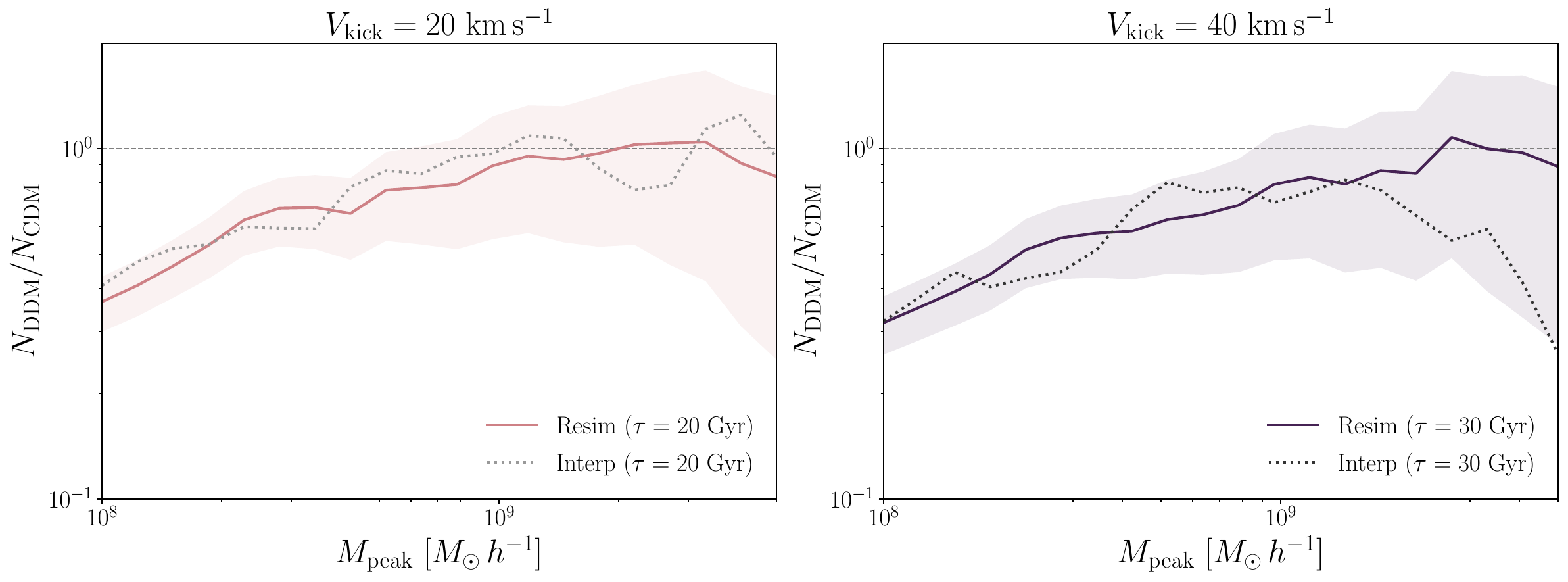}
\caption{Average SHMF suppression at $z=0$ for the MW-like host halos in our MW-like resimulations (described in Appendix \ref{sec:appendix_a}) measured directly from the resimulations (solid lines) and using the interpolating procedure described in Appendix \ref{sec:appendix_b}. Results are shown for the resimulated DDM models with $\tau = 20\Gyr$, $V_{\mathrm{kick}} = 20\kms$ (left panel) and $\tau = 30\Gyr$, $V_{\mathrm{kick}} = 40\kms$ (right panel) as a function of peak subhalo virial mass $M_{\mathrm{peak}}$. All SHMFs are restricted to subhalos above a conservative resolution threshold of $V_{\rm peak} > 10 \kms$ and $V_{\rm max} > 9 \kms$. Shaded bands indicate $68\%$ confidence interval Poisson uncertainties on the simulation measurements; the SHMF suppression predicted by our interpolating procedure based on the \citet{Wang:2014} simulations is consistent with that measured from our MW-like resimulations at the $\sim 1\sigma$ level.
}
\label{fig:interp-resim}
\end{figure*}

\begin{figure*}
\includegraphics[width=\textwidth]{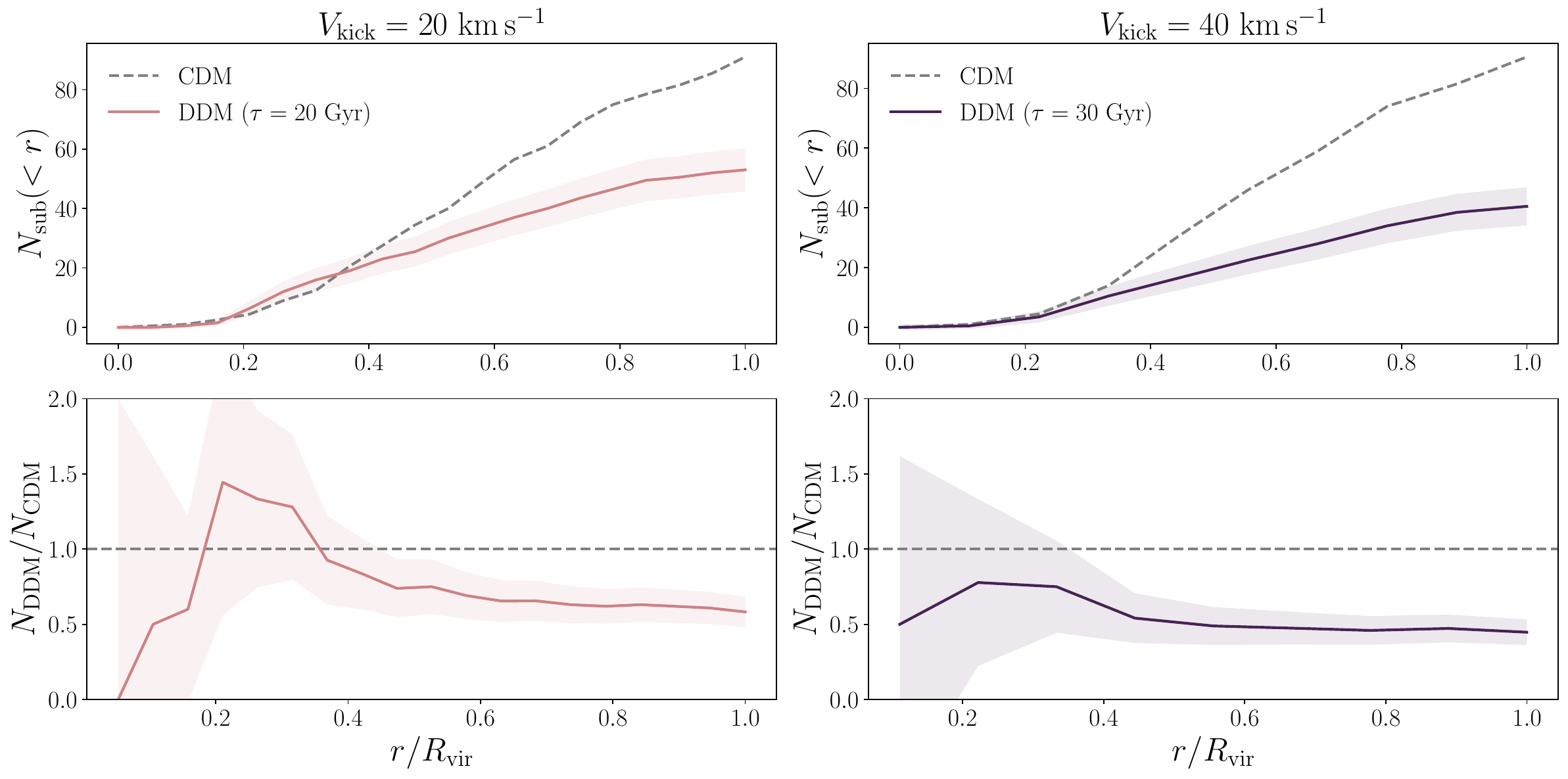}
\caption{Average subhalo radial distributions at $z=0$ as a function of distance from the center of the host halo for the CDM (dashed) and DDM (solid) resimulations of MW-like halos described in Appendix \ref{sec:appendix_a}. Results are shown for the resimulated DDM models with $\tau = 20\Gyr$, $V_{\mathrm{kick}} = 20\kms$ (red) and $\tau = 30\Gyr$, $V_{\mathrm{kick}} = 40\kms$ (purple). The bottom panels show the ratio of the radial distribution in each DDM model relative to CDM. All results are restricted to subhalos above a conservative resolution threshold of $V_{\rm peak} > 10 \kms$ and $V_{\rm max} > 9 \kms$. Shorter DDM decay lifetimes result in fewer surviving subhalos relative to CDM. This effect is more pronounced for models with higher kick velocities and does not strongly depend on radius from the center of the host halo. Shaded bands indicate $68\%$ confidence interval Poisson uncertainties on the simulation measurements.
}
\label{fig:shrf-resim}
\end{figure*}

As shown in \figref{interp-resim}, the DDM SHMF provided by the interpolation function also matches that derived from our MW-like resimulations at the $\sim 1\sigma$ level.
We note that, while one of these resimulations covers the same point in $\tau$, $V_{\rm kick}$ space as one of the \citet{Wang:2014} simulations, the second does not; thus, the interpolating function is used to compare with our MW-like resimulations in both cases.
Combining the results from \figref{sim-interp} and \figref{interp-resim} demonstrates that the DDM SHMF suppression derived and adopted in our fiducial analysis based on the \citet{Wang:2014} simulations is consistent with that from simulations of systems specifically chosen to resemble the MW, lending confidence to our constraints.
Finally, we validate the behavior of the subhalo radial distribution of the \citet{Wang:2014} simulations (\figref{shrf}) by comparing to the distribution for our MW-like simulations in \figref{shrf-resim}.

\section{Evolution of the DDM Subhalo and Halo Mass Functions}
\label{sec:appendix_c}

\begin{figure*}
\includegraphics[width=\textwidth]{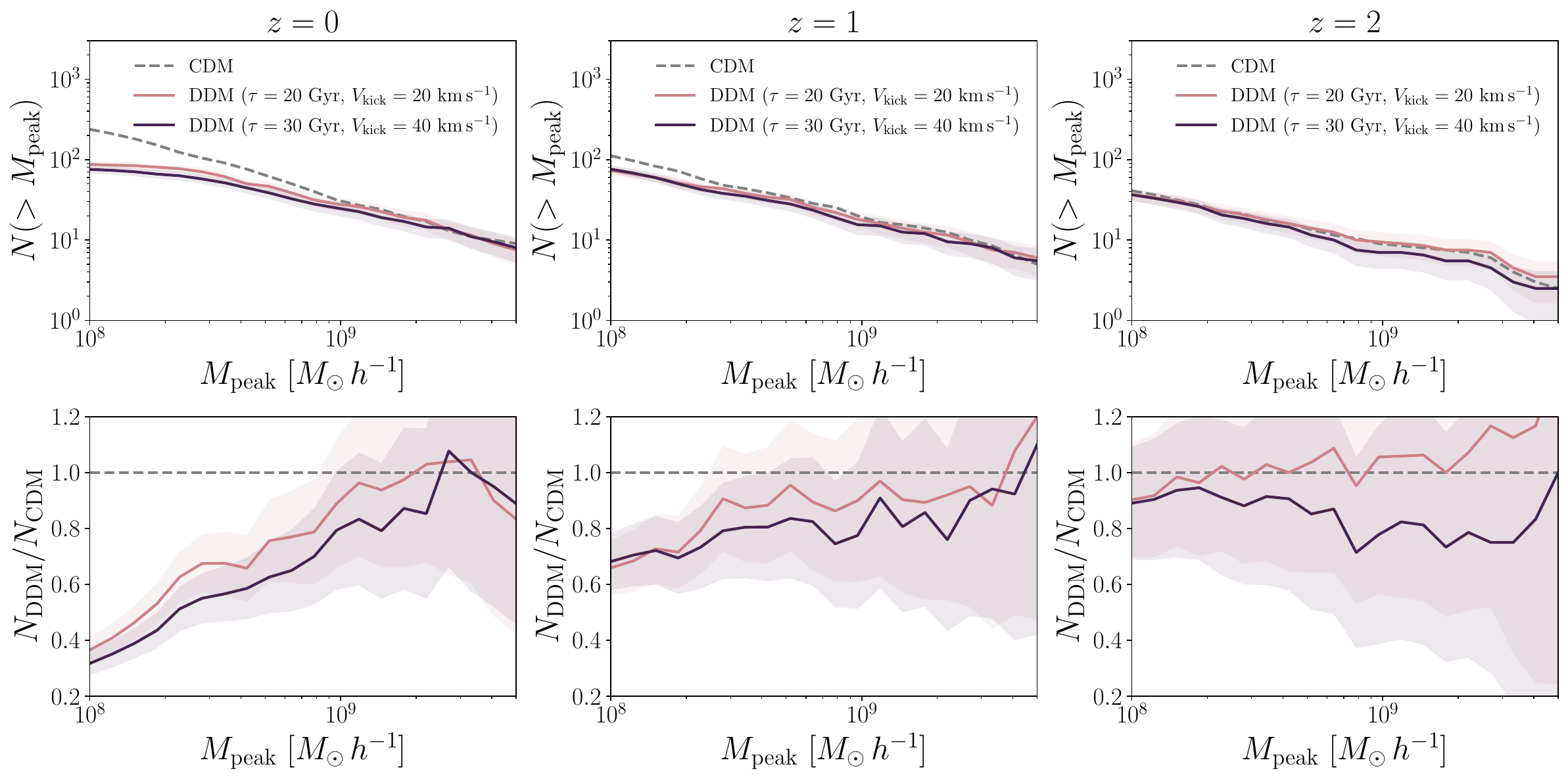}
\caption{Evolution of the subhalo $M_{\rm peak}$ functions for the CDM and DDM resimulations of MW-like halos described in Appendix \ref{sec:appendix_a}. Results are shown at $z=0$ (left panel), $z=1$ (middle panel), and $z=2$ (right panel), and the bottom panels show the corresponding SHMF suppression. Results are shown for the resimulated DDM models with $\tau = 20\Gyr$, $V_{\mathrm{kick}} = 20\kms$ (red) and $\tau = 30\Gyr$, $V_{\mathrm{kick}} = 40\kms$ (purple). All SHMFs are restricted to subhalos above a conservative resolution threshold of $V_{\rm peak} > 10 \kms$ and $V_{\rm max} > 9 \kms$; note that $V_{\mathrm{peak}}$ is typically achieved at much earlier times than shown here ($z\sim 4$). The suppression of the DDM SHMF only sets in significantly at late times, consistent with the intuition developed in \secref{analytic_disruption}. Shaded bands indicate $68\%$ confidence interval Poisson uncertainties on the simulation measurements.
}
\label{fig:redshift}
\end{figure*}

We use the DDM resimulations of MW-like systems presented in Appendix \ref{sec:appendix_a} to study the evolution of the DDM subhalo and halo mass functions. In particular, \figref{redshift} shows the evolution of the average SHMF for subhalos of the two MW-like hosts at $z = 0$, 1, and 2. The DDM SHMFs are consistent with the corresponding CDM SHMFs for $z\gtrsim 2$, accounting for Poisson uncertainties, and only become significantly suppressed at later times. This reflects the combined effects of mass loss due to DM decays, which can push subhalos below the mass resolution limit of our simulations (\secref{analytic_disruption}), and the enhanced tidal disruption of these systems with reduced central densities relative to CDM. This late-time suppression differentiates DDM from WDM-like models that suppress the linear matter power spectrum.

\begin{figure*}
\includegraphics[width=\textwidth]{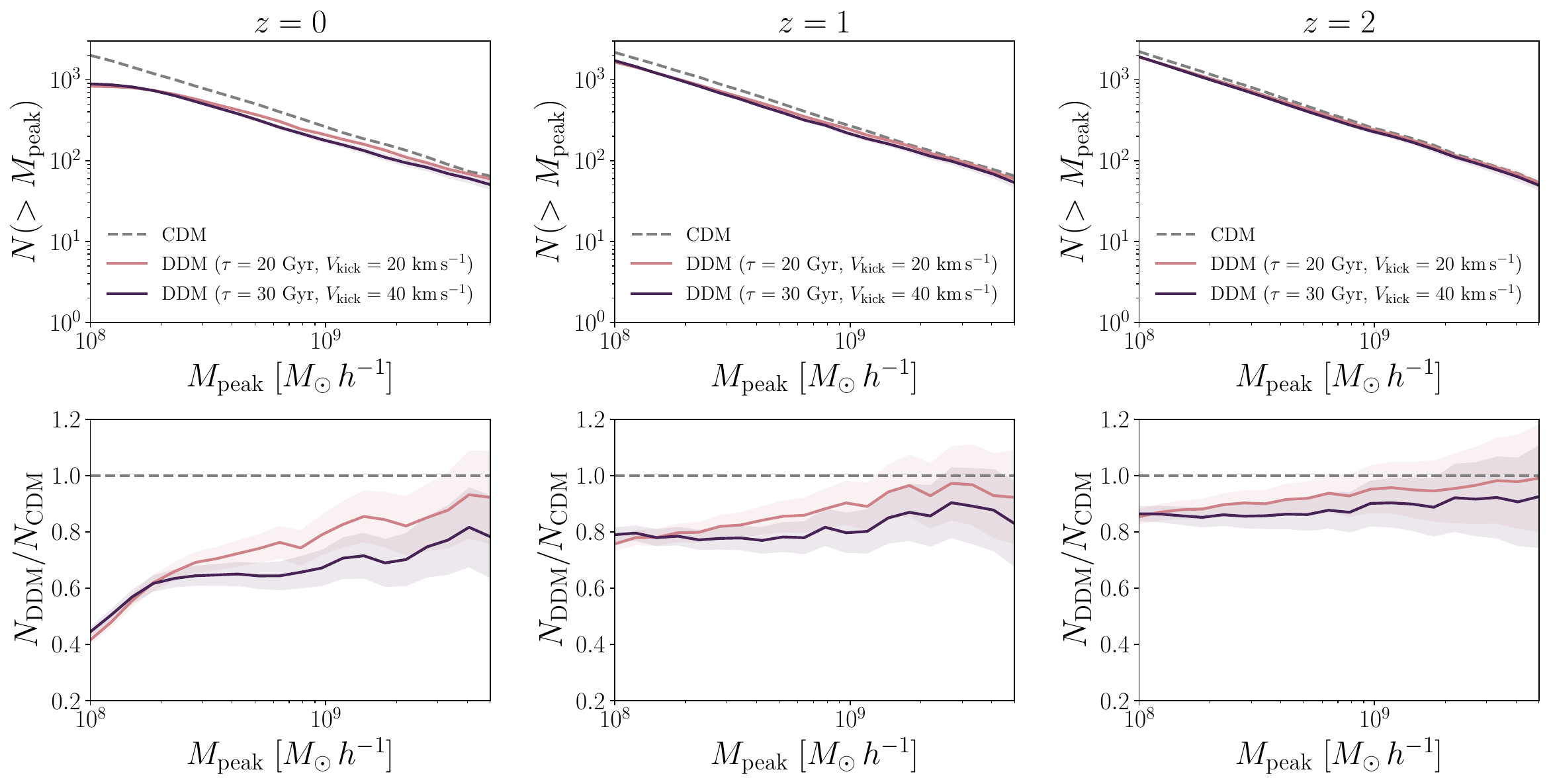}
\caption{Evolution of the \emph{isolated} halo $M_{\rm peak}$ functions for the CDM and DDM resimulations of MW-like halos described in Appendix \ref{sec:appendix_a}. These measurements are presented analogously to \figref{redshift}. The abundance of isolated halos is significantly suppressed at late times in DDM and is slightly less severe than the suppression of subhalo abundances, consistent with mass loss due to DM decays driving this effect. Shaded bands indicate $68\%$ confidence interval Poisson uncertainties on the simulation measurements.}
\label{fig:redshift-isolated}
\end{figure*}

Finally, \figref{redshift-isolated} shows peak velocity functions at $z = 0$, 1, and 2 for \emph{isolated} halos surrounding our resimulated MW-like systems, most of which lie within a $\sim 3\Mpc$ radius from the center of the host halo that contains $\sim 90\%$ of the highest-resolution particles \citep{Wang210211876}.\footnote{In particular, we analyze isolated halos by choosing systems with \textsc{ROCKSTAR} \texttt{upid} equal to $-1$.} Like the SHMF, the suppression of the isolated DDM halo mass function only sets in significantly at late times. The suppression of isolated halo abundances is only slightly less severe than that for subhalos, consistent with mass loss due to DM decays driving the disruption. This effect---i.e., severe mass loss for isolated halos---differentiates DDM from models like self-interacting DM in which late-time physics preferentially disrupts subhalos at late times (e.g., \citealt{Tulin170502358,Nadler200108754}).

\section{Galaxy--Halo Connection Model and Parameters}
\label{sec:appendix_d}

\begin{table*}[ht!]
    \caption{Galaxy--Halo Connection Model Parameters.}
    \begin{tabular*}{\textwidth}{r | l @{\extracolsep{\fill}} c c}
        \hline\hline
        Parameter & Physical Interpretation & Symbol & Units \\
        \hline
        Faint-end slope & Power-law slope of satellite luminosity function & $\alpha$ & none \\
        Luminosity scatter & Scatter in luminosity at fixed $V_{\rm peak}$ & $\sigma_M$ & dex \\
        50\% occupation mass & Mass at which 50\% of halos host galaxies & $\mathcal{M}_{50}$ & $\Msun$ \\
        Baryonic effects & Strength of subhalo disruption due to baryons & $\mathcal{B}$ & none \\
        Occupation scatter & Scatter in galaxy occupation fraction & $\sigma_{\mathrm{gal}}$ & dex \\
        Size amplitude & Amplitude of galaxy--halo size relation & $\mathcal{A}$ & $\mathrm{pc}$ \\
        Size scatter & Scatter in half-light radius at fixed halo size & $\sigma_{\log R}$ & dex \\
        Size power-law index & Power-law index of galaxy--halo size relation & $n$ & none \\
        Decay lifetime & Inverse of the DM particle decay rate & $\tau$ & $\mathrm{Gyr}$ \\
        \hline
    \end{tabular*}
    \label{tab:model}
\end{table*}

We follow \citet{PaperII} in modeling the galaxy--halo connection used in our forward model (\secref{fitting}) with eight free parameters and introduce one additional parameter for DDM particle lifetime.
These parameters are summarized in \tabref{model}.

\subsection{Satellite Luminosities}

We employ an abundance-matching procedure that relates the absolute $V$-band magnitude of satellites, $M_V$ to the peak circular velocity of subhalos, $V_\text{peak}$ \citep{Nadler180905542}.
This relation is extended to dim satellites by allowing the faint-end slope of the satellite luminosity function, $\alpha$, and the lognormal scatter in luminosity at fixed $V_\text{peak}$, $\sigma_M$, to be free parameters.
While this abundance-matching model does not capture the entire star formation histories of ultra-faint dwarf galaxies, it is consistent with current MW satellite data \citep{PaperI}.

\subsection{Satellite Sizes}

The mean predicted size of each satellite at accretion is set according to
\begin{equation}
    r_{1/2} \equiv \mathcal{A} \left(\frac{R_\text{vir}}{R_0}\right)^n,
    \label{eq:size}
\end{equation}
where $\mathcal{A}$ and $n$ are free parameters corresponding to the amplitude and power-law index of the galaxy--halo size relation, respectively, and $R_\text{vir}$ denotes the subhalo virial radius as measured at accretion; $R_0 = 10\kpc$ is a normalization constant.

In our inference, satellite sizes are drawn from a lognormal distribution with mean given by \eqnref{size} and standard deviation $\sigma_{\log R}$, a further free parameter.
While post-infall effects---including adiabatic decays---can shrink or enlarge satellites, \citet{PaperII} found that our results are not sensitive to these effects while using a model for satellite size evolution due to tidal stripping; thus, these effects are not modeled here.

\subsection{Subhalo Disruption due to Baryonic Effects}

We incorporate the effects of baryonic physics---particularly the tidal influence on the Galactic disk---on our simulated subhalo populations following \citet{Garrison-Kimmel170103792} and \citet{Nadler171204467}.
The strength of the disruption is modeled using the free parameter $\mathcal{B}$ for which $\mathcal{B} = 1$ corresponds to fiducial hydrodynamical predictions \citep{Nadler171204467} and larger (smaller) values of $\mathcal{B}$ correspond to more (less) effective subhalo disruption.
For each subhalo, we set
\begin{equation}
    p_\text{disrupt} \equiv (p_{\text{disrupt}, 0})^{1/\mathcal{B}},
\end{equation}
where $p_{\text{disrupt}, 0}$ is the fiducial disruption probability given by the random forest algorithm of \citet{Nadler171204467}.

\subsection{Galaxy Formation Efficiency}

We parameterize the fraction of halos that host galaxies of any mass---the \emph{galaxy occupation fraction}---following \citet{10.1093/mnras/stz1992},
\begin{equation}
    f_\text{gal}(\mathcal{M}_\text{peak}) \equiv \frac{1}{2} \left[1 + \operatorname{erf}\left(\frac{\mathcal{M}_\text{peak} - \mathcal{M}_{50}}{\sqrt{2} \sigma_\text{gal}}\right)\right],
\end{equation}
where $\mathcal{M}_\text{peak}$ is the largest virial mass a subhalo ever attains, $\mathcal{M}_{50}$ is the peak halo mass at which 50\% of halos host galaxies of any mass, and $\sigma_\text{gal}$ is the width of the galaxy occupation fraction; in our inference, $\mathcal{M}_{50}$ and $\sigma_\text{gal}$ are free parameters.

\subsection{Orphan Satellites}

We account for orphan satellites---subhalos that have been artificially disrupted by approaching the resolution limit of our simulations---by following the prescription of \citet{Nadler180905542}, which identifies disrupted subhalos in each simulation, interpolations their orbits to $z=0$ using a softened gravitational force law and a dynamical friction model, and accounts for tidal stripping with a mass-loss model.
The effective abundance of orphans is parameterized by setting their disruption probabilities equal to
\begin{equation}
    p_\text{disrupt} \equiv (1 - a_\text{acc})^{\mathcal{O}},
\end{equation}
where $a_\text{acc}$ is the final scale factor at which each subhalo enters the virial radius of the MW analog in the simulations, and $\mathcal{O}$ captures deviations from disruption probabilities in hydrodynamic simulations.
We follow \citet{Nadler180905542} by fixing $\mathcal{O} = 1$; note that \citet{PaperII} found that the results of a CDM fit to the observed MW satellite population are insensitive to the value of $\mathcal{O}$.


\end{document}